\documentclass[lettersize,journal]{IEEEtran}
\usepackage{stmaryrd}
\usepackage{amsmath}
\usepackage{amssymb}
\usepackage{psfrag}
\usepackage{graphicx}
\usepackage{epstopdf}
\usepackage{multirow}
\usepackage{booktabs}
\usepackage{siunitx}
\usepackage{cite}
\usepackage{algorithmic}
\usepackage{algorithm}
\usepackage{svg}
\usepackage{mathtools}
\usepackage{dutchcal}
\usepackage{hyperref}
\usepackage{bm}

\DeclareMathOperator*{\argmin}{argmin}

\usepackage{float}
\ifCLASSOPTIONcompsoc
\usepackage[caption=false,font=normalsize,labelfont=sf,textfont=sf]{subfig}
\else
\usepackage[caption=false,font=footnotesize]{subfig}
\fi

\usepackage{flushend}

\begin{document}

\title{2D-RC: Two-Dimensional Neural Network Approach for OTFS Symbol Detection}
\author{Jiarui Xu, Karim Said, Lizhong Zheng, and Lingjia Liu
\thanks{J. Xu, K. Said, and L. Liu are with Wireless@Virginia Tech, the Bradley Dept. of ECE at Virginia Tech. L. Zheng is with the EECS Department at the Massachusetts Institute of Technology. 
}%
}
%



\maketitle

\begin{abstract}
Orthogonal time frequency space (OTFS) is a promising modulation scheme for wireless communication in high-mobility scenarios.
Recently, a reservoir computing (RC) based approach has been introduced for online subframe-based symbol detection in the OTFS system, where only a limited number of over-the-air (OTA) pilot symbols are utilized for training.
However, this approach does not leverage the domain knowledge specific to the OTFS system to fully unlock the potential of RC.
This paper introduces a novel two-dimensional RC (2D-RC) method that incorporates the domain knowledge of the OTFS system into the design for symbol detection in an online subframe-based manner.
Specifically, as the channel interaction in the delay-Doppler (DD) domain is a two-dimensional (2D) circular operation, the 2D-RC is designed to have the 2D circular padding procedure and the 2D filtering structure to embed this knowledge.
With the introduced architecture, 2D-RC can operate in the DD domain with only a single neural network, instead of necessitating multiple RCs to track channel variations in the time domain as in previous work.
Numerical experiments demonstrate the advantages of the 2D-RC approach over the previous RC-based approach and compared model-based methods across different OTFS system variants and modulation orders.

\end{abstract}


%
\IEEEpeerreviewmaketitle


\section{Introduction}

Next-generation wireless communication systems are required to support reliable communication quality in high-speed scenarios, such as high-speed railways, unmanned aerial vehicles, and low earth orbit~\cite{series2015imt}.
However, in such scenarios, orthogonal frequency division multiplexing (OFDM), which is a key physical layer waveform of 4G LTE-Advanced and 5G NR~\cite{5GAI}, suffers from the inter-carrier interference (ICI) caused by the high Doppler spread.
Recently, OTFS modulation has emerged as a promising modulation scheme for reliable communications in high-mobility scenarios~\cite{hadani2017orthogonal}.
Different from OFDM which multiplexes information symbols in the time-frequency (TF) domain, OTFS is a 2D modulation scheme that transmits information symbols in the DD domain.
In the DD domain, each transmitted symbol spreads over the TF domain and experiences the full TF-domain channel.
Therefore, OTFS provides the potential of achieving full channel diversity~\cite{wei2021orthogonal, yuan2023new}.
More recent work has also analyzed the channel predictability in the DD domain of the OTFS system~\cite{mohammed2022otfs}.



The benefits of adopting OTFS modulation in high-mobility scenarios have attracted substantial interest in investigating low-complexity equalization techniques for the OTFS system.
Existing approaches can be roughly divided into two branches: model-based methods and learning-based approaches.
Model-based approaches are designed based on analyzing the input-output relationship and the structure of the equivalent channel matrix in the OTFS system.
Specifically, a set of linear equalizers~\cite{surabhi2019low, tiwari2019low, zou2021low} is introduced to conduct low-complexity linear minimum mean square error (LMMSE) detection by taking advantage of the channel structure.
For example, the double block circulant structure of the channel in the DD domain is leveraged in~\cite{surabhi2019low} under the bi-orthogonal pulse shaping assumption.
The quasi-banded structure of the time-domain equalization matrix is utilized for low complexity matrix inversion in~\cite{tiwari2019low}.
The block circulant structure of the DD-domain equivalent channel in the OFDM-based OTFS system with rectangular pulse shaping is exploited in~\cite{zou2021low}.

Furthermore, multiple non-linear detectors are developed to approach the maximum \emph{a posteriori} (MAP) performance with a lower complexity than the MAP~\cite{raviteja2018interference, yuan2021iterative, liu2021message, thaj2020low, zhang2021low, yuan2020simple, qu2021TCOMLowcomplexSD, shan2022orthogonal, li2021cross, li2021hybridpic}.
For instance, the message passing algorithm (MPA)~\cite{raviteja2018interference} is developed to conduct the low complexity detection based on the Gaussian assumption of the interference and the sparsity of the channel matrix in the DD domain.
The hybrid MAP and parallel interference cancellation (Hybrid-MAP-PIC) algorithm in~\cite{li2021hybridpic} combines the symbol-wise MAP approach with the MPA to achieve a better performance than the MPA at the cost of a higher computational complexity.
When it comes to the case with factional Doppler shifts,  the cross-domain iterative detection approach in~\cite{li2021cross} is designed to iteratively perform the LMMSE detection in the time domain and the symbol-by-symbol detection in the DD domain.
The cross-domain method can approach the performance of the symbol-wise MAP detector.
However, the computational complexity of this algorithm is in a cubic order of the subframe size, which is computationally expensive for practical systems.
In the continuous-Doppler-spread scenario, the iterative least squares minimum residual (LSMR)-based equalizer~\cite{qu2021TCOMLowcomplexSD} is introduced to iteratively conduct LSMR detection and interference cancellation.
The LSMR-based approach outperforms MPA and maintains a lower computational complexity than the MPA.
While model-based approaches are explainable and easy to analyze, they usually rely on explicit system modeling and accurate channel state information (CSI) estimation.
The performance of such methods suffers from system model mismatch and channel estimation error.
Learning-based detection approaches leverage the power of neural networks (NNs) to learn the mapping from the received signal to the transmitted one, which does not necessarily require explicit system modeling and knowledge of the CSI.
Existing learning-based algorithms can be broadly classified as offline learning methods and online learning methods.
Most existing learning-based techniques are offline learning methods, which rely on extensive offline training data and a long training time~\cite{enku2021two, naikoti2021low, enku2022deep, zhang2022gaussian, zhang2023graph}.
Approaches under this category, such as convolutional neural network (CNN)-based techniques in~\cite{enku2021two, enku2022deep}, the multi-layer perceptron (MLP)-based method in~\cite{naikoti2021low}, the GAMP-NET in~\cite{zhang2022gaussian}, and the graph neural network (GNN)-based algorithm in~\cite{zhang2023graph}, train NNs offline with a large amount of training data and then directly deploy the trained NN online.
However, when the online data distribution is different from offline training data distribution, these offline learning approaches may have the ``uncertainty in generalization” issue~\cite{shafin2020artificial} and experience performance degradation.
Furthermore, due to the dynamic channel environment, the modern cellular system has dynamic transmission modes through rank adaptation, link adaptation, and scheduling operations, which are all performed on a subframe basis~\cite{4GMIMO_OFDM}. 
The discrepancy between the mode of offline training and online deployment may prevent the offline-trained models from being adopted online.
To address the above challenges, an online learning algorithm for the OTFS symbol detection is developed in our previous work~\cite{zhou2022learningotfs}, which can be learned with only the limited over-the-air (OTA) training pilots and dynamically updated on a subframe basis. 
This approach utilizes reservoir computing (RC), which is a particular type of recurrent neural network (RNN), to achieve online subframe-based learning.
Compared with a typical RNN, RC only contains a few trainable parameters, allowing for an efficient and simple training procedure with limited training data.
While the previous RC-based approach can achieve compelling performance, it operates in the time domain and therefore requires multiple RCs to track the channel changes.
Furthermore, it directly applies the RC structure in~\cite{zhou2019}, which is designed for the OFDM system, and does not incorporate the domain knowledge of the OTFS system to unleash the full potential of RC.

In this work, we introduce a novel 2D-RC structure for the online subframe-based symbol detection task in the OTFS system.
The introduced 2D-RC retains the advantages of RC that can be learned with limited OTA training pilots within each subframe and dynamically updated on a subframe basis, which differentiates it from existing offline learning methods that rely on extensive training data and a long training time.
Compared with the RC-based online learning method in~\cite{zhou2022learningotfs}, it further incorporates the domain knowledge of the OTFS system into the design. 
Specifically, the channel in the DD domain works as a 2D circular operation over the transmitted symbols in the OTFS system.
This domain knowledge is integrated into the 2D-RC through the design of the 2D circular padding operation and the 2D filtering structure. 
By incorporating the domain knowledge, 2D-RC can operate in the DD domain with only a single NN, which is shown to be more effective than the previous RC-based approach with multiple RCs in the time domain.
The contributions of this work are summarized as the following:
\begin{itemize}
    \item 
    We introduce a novel 2D-RC structure to conduct symbol detection in the OTFS system in an online subframe-based fashion.
    The 2D-RC embeds the domain knowledge of the 2D circular channel interaction in the DD domain into its design, which has the 2D circular padding and the 2D filtering structure.
    By embedding the domain knowledge, 2D-RC can achieve substantial performance improvement over the previous RC-based approach in different variants of the OTFS system and under different modulation orders.
    \item The 2D-RC approach offers better generalization ability than the previous RC-based approach.
    Instead of requiring multiple RCs to achieve a satisfactory performance, the 2D-RC necessitates only a single NN for processing, which eliminates the requirement to configure the number of RCs.
    Evaluation results show that 2D-RC with a single NN achieves better performance than the multiple-RC approach in various compared scenarios.
    \item 
    The 2D-RC can be readily adapted to different variants of the OTFS system without the requirement of channel knowledge, which is different from model-based approaches that require knowledge of the CSI and are tailored for specific types of OTFS systems with specific assumptions.
    Experimental results reveal the advantages of the 2D-RC over the compared model-based approaches across different OTFS system variants.
\end{itemize}

The remainder of this paper is organized as follows.
Sec.~\ref{sec:prelim} briefly discusses the preliminaries of RC.
Sec.~\ref{sec:sys_model} presents the basics of the OTFS system.
Sec.~\ref{sec:introduced_approach} introduces the designed 2D-RC approach.
Sec.~\ref{sec:complexity} analyzes the complexity of 2D-RC.
Sec.~\ref{sec:experiments} provides the performance evaluation of the 2D-RC with our previous RC-based approach and model-based detection methods for the OTFS system. 
The paper is concluded in Sec.~\ref{sec:conclusion}.

\begin{figure*}
\centering
\includegraphics[width=0.7\linewidth]{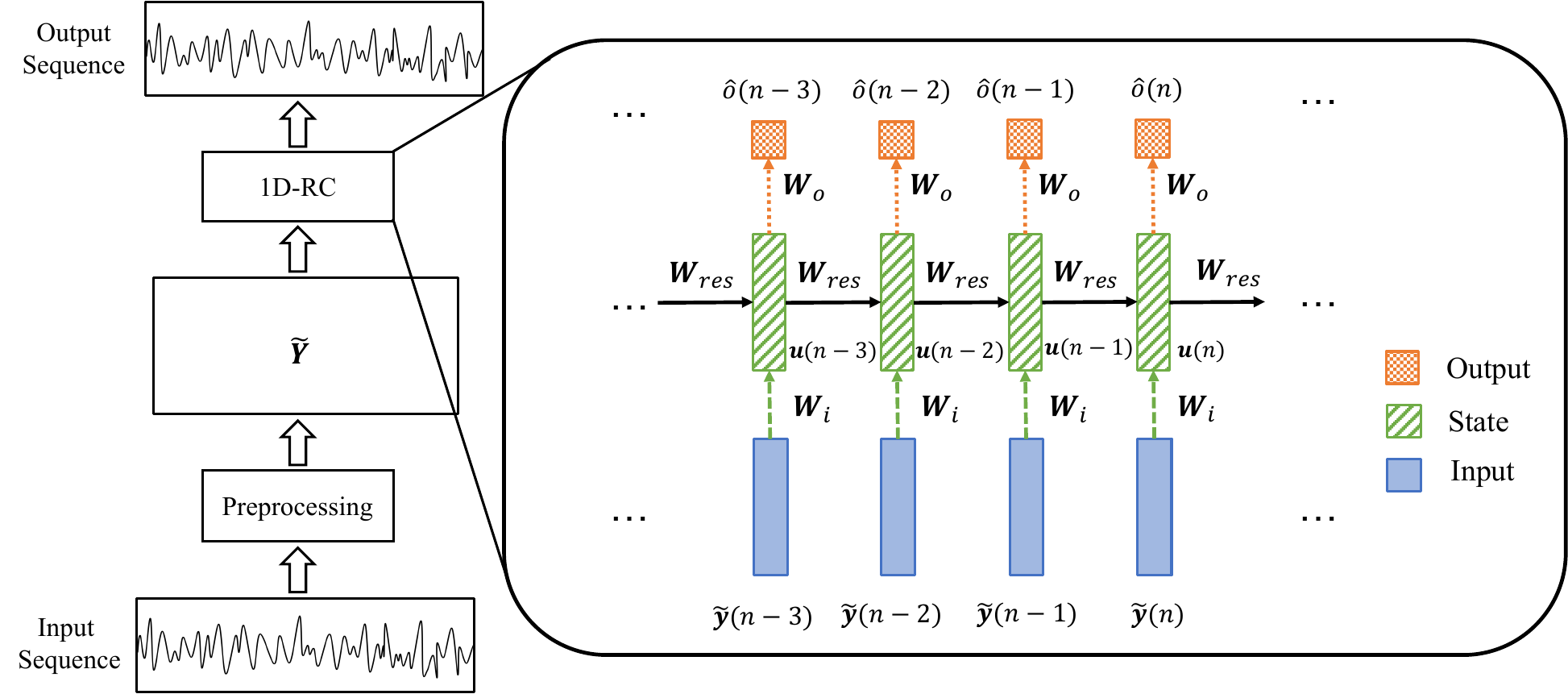}
\caption{1D-RC Structure. For simplicity, the extended state and nonlinear function are ignored here. In the figure, the target output is a sequence with $N_o=1$. 
}
\label{figs:1d_rc_structure}
\end{figure*}

\begin{figure}
\centering
\includegraphics[width=0.7\linewidth]{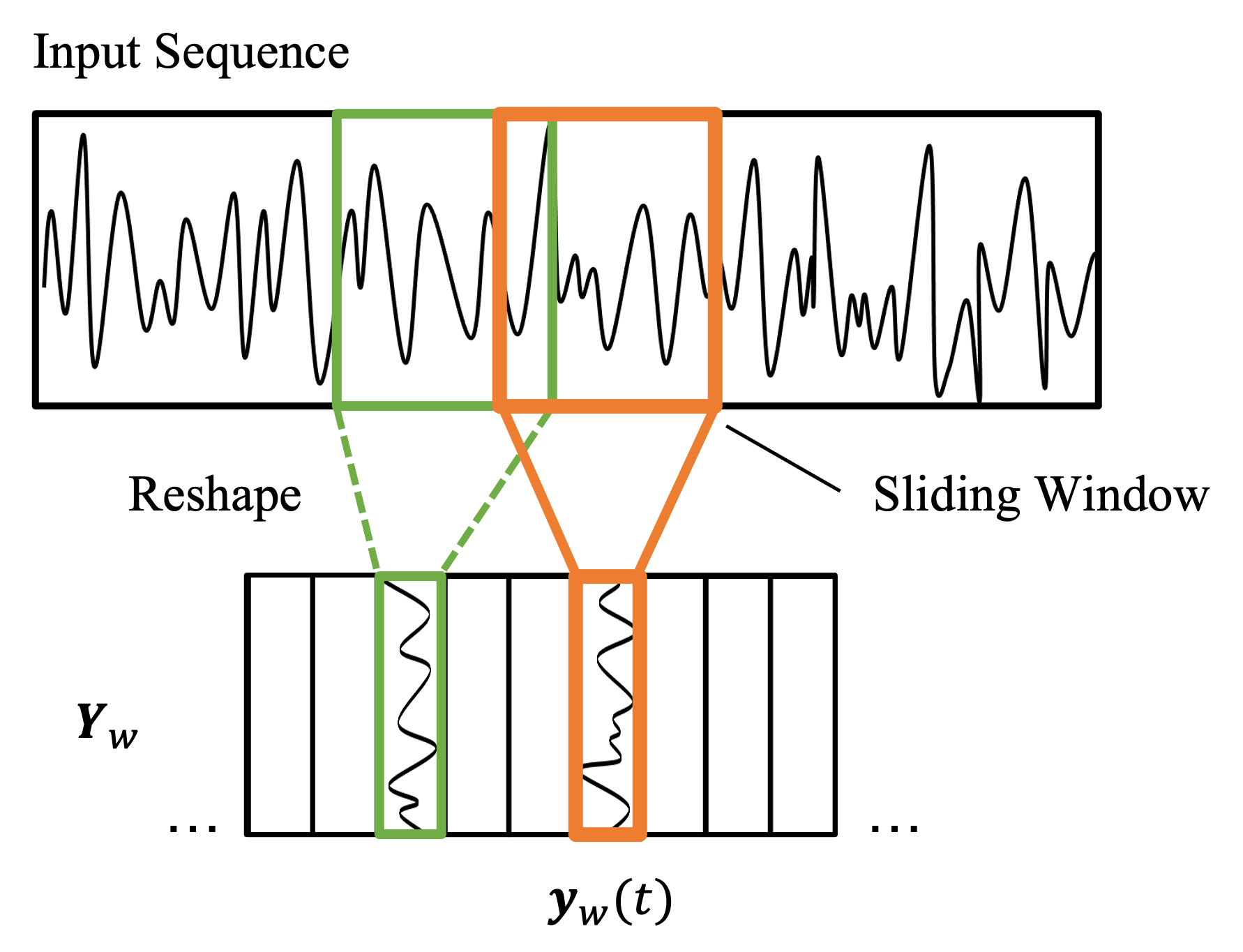}
\caption{The windowing process in 1D-RC.}
\label{figs:1d_rc_pre_process}
\end{figure}

\textbf{Notations}: 
Non-bold letter, bold lowercase letter, bold uppercase letter, and bold Euler script letter, i.e., $x$, $\bm{x}$, $\bm{X}$, and $\mathbcal{X}$, denote scalar, vector, matrix, and tensor, respectively.
$\mathbb{C}$ represents the complex number set and $\mathbb{R}$ is the real number set.
$\boldsymbol{F}_M$ and $\boldsymbol{F}_M^H$ denote the normalized $M$-point discrete Fourier transform (DFT) and $M$-point inverse discrete Fourier transform (IDFT), respectively. 
$(\cdot)^\dag$ represents the Moore–Penrose inverse.
$(\cdot)^T$ denotes the transpose operation.
$\langle\cdot\rangle_M$ and $\lfloor\cdot\rfloor$ stand for the modulo operator of divider $M$ and the floor operation, respectively.
$\mathrm{vec}(\cdot)$ denotes the operation of vectoring the matrix by stacking along the columns, and $\mathrm{vec}^{-1}(\cdot)$ denotes unfolding the vector to a matrix by filling the matrix column by column.
$\bm{I}_{M}$ is a $M$-dimensional identity matrix.
$\odot$ denotes the Hadamard product operation between two matrices.
The $n$-mode Hadamard product between the matrix $\bm{U} \in \mathbb{C}^{I_n\times I_{n+1}}$ and the $N$-dimensional tensor $\mathbcal{X} \in \mathbb{C}^{I_1\times I_2 \times \dots \times I_N}$ is defined as
\begin{multline*}
    (\bm{U} \odot_n \mathbcal{X})[i_1, \dots,i_{n}, i_{n+1}, \dots, i_N] \\
    = U[i_{n}, i_{n+1}]\cdot X[i_1, \dots,i_{n}, i_{n+1}, \dots, i_N],
\end{multline*}
where $U[i_{n}, i_{n+1}]$ is the $(i_{n}, i_{n+1})$-th element in $\bm{U}$, and $X[i_1, \dots,i_{n}, i_{n+1}, \dots, i_N]$ is the $(i_1, \dots,i_{n}, i_{n+1}, \dots, i_N)$-th element in $\mathbcal{X}$.
The concatenation of two tensors $\mathbcal{X}_1$ and $\mathbcal{X}_2$ along the $n$-th dimension is represented by $\text{cat}_n(\mathbcal{X}_1, \mathbcal{X}_2)$.

\section{Preliminaries -- Reservoir computing}
\label{sec:prelim}

RC is a class of RNNs for processing temporal or sequential data.
It consists of an RNN-based reservoir to map inputs into a high-dimensional state space and an output layer to learn the projection of the target to the high-dimensional state space~\cite{tanaka2019recent}.
The characteristic feature of RC is that the reservoir weights are fixed after being randomly initialized and only the output layer is updated through a simple linear regression.
The fast and simple training process differentiates RC from other RNNs and enables its broad application in different research areas~\cite{jalalvand2015real, tong2018reservoir,triefenbach2010phoneme, verstraeten2006reservoir}.
Recently, RC has shown its effectiveness in the symbol detection task for both the OFDM system~\cite{mosleh2017brain, zhou2019, zhou2020rcnet, xu2021rcstruct, xu2023DetectToLearn, li2023mmstructnet} and the OTFS system~\cite{zhou2022learningotfs}.
In this work, we focus on customizing RC for the symbol detection task in the OTFS system, instead of directly applying the existing structure of RC as in~\cite{zhou2022learningotfs}.
Before we introduce our designed RC structure, we briefly review the processing procedures of RC that have been adopted in previous works~\cite{zhou2019, zhou2020rcnet, xu2021rcstruct, xu2023DetectToLearn, zhou2022learningotfs}.
For ease of discussion, we refer to the existing RC structure as ``1D-RC" for the remainder of this paper.

\subsection{Pre-processing}
\label{sec:prelim_pre_process}

\subsubsection{Windowing}
Suppose the sequential input is $\bm{Y} \triangleq [\bm{y}(0), \bm{y}(1),\dots,\bm{y}(L_t-1)] \in \mathbb{C}^{N_y \times L_t}$, where $N_y$ is the input dimension, and $L_t$ is the sequential length of the input.
A sliding window is adopted in the pre-processing procedure to increase the short-term memory of RC~\cite{zhou2019}.
Specifically, the windowed input is obtained by stacking a sequence of input vectors within the sliding window length $N_w$, which can be written as ${\bm{y}}_w(t) \triangleq [\bm{y}(t)^T, \bm{y}(t-1)^T, \dots, \bm{y}(t-N_w+1)^T]^T$.
The ${\bm{y}}_w(t) \in \mathbb{C}^{N_i}$ is the windowed input vector at time step $t$ ($t = 0, 1, \dots, L_t-1$), where $N_i=N_yN_w$ is dimension of the windowed input.
When $t<N_w-1$, zeros are added at the end of ${\bm{y}}_w(t)$ to maintain the input length of $N_i$.
The matrix form of the windowed input is obtained by concatenating the windowed input vector at each time step, i.e., $\bm{Y}_w \triangleq [\bm{y}_w(0), \bm{y}_w(1),\dots,\bm{y}_w(T-1)] \in \mathbb{C}^{N_i \times L_t}$.
The windowing process is illustrated in Fig.~\ref{figs:1d_rc_pre_process}.
For simplicity, $N_y$ is assumed to be $1$ in the figure.

\subsubsection{Padding}
RC requires a degree of forgetfulness to remove the impact from the random initialization of the internal state~\cite{lukovsevivcius2009reservoir}.
Therefore, the input is further padded with zeros at the end to facilitate the learning process of the optimal forget length for the internal state.
The padded input is denoted as $\tilde{\bm{Y}} \triangleq [\bm{Y}_w, \bm{0}_{N_i\times L_f}] \in \mathbb{C}^{N_i \times (L_t+L_f)}$, where $L_f$ is the maximum forget length of the internal state and $\bm{0}_{N_i\times L_f}$ is a zero matrix of size $N_i\times L_f$.

\begin{figure*}[ht]
\centering
\includegraphics[width=0.8\linewidth]{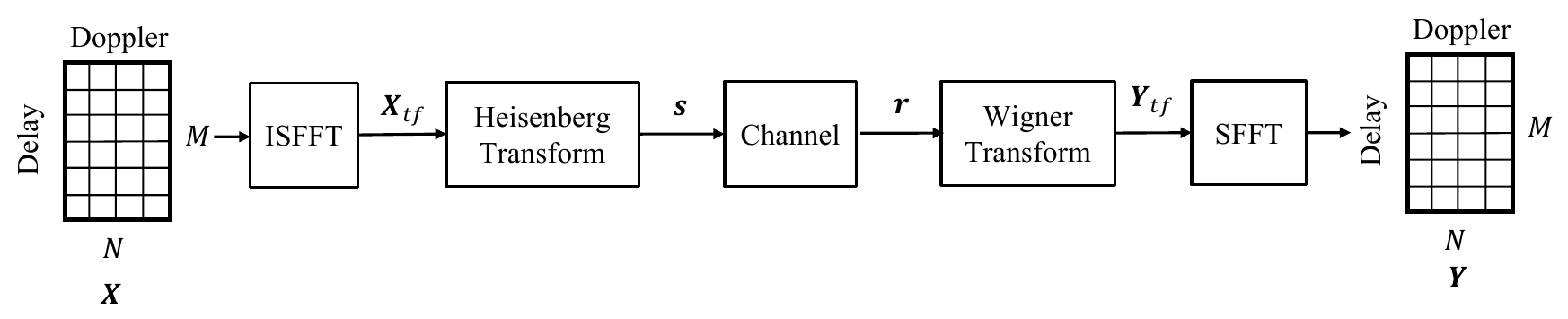}
\caption{OTFS system diagram.}
\label{fig:otfs_system}
\end{figure*}

\subsection{Structure of 1D-RC}
As shown in Fig.~\ref{figs:1d_rc_structure}, the 1D-RC has a recurrent structure as RNNs.
Denote $\tilde{\bm{y}}(n) \in \mathbb{C}^{N_i}$ as the $n$-th column of $\tilde{\bm{Y}}$ ($n=0, 1, \dots, L_t+L_f-1$).
The state transition equation of 1D-RC is expressed as
\begin{align}
    \bm{u}(n) = f(\bm{W}_i\;\tilde{\bm{y}}(n) + \bm{W}_{\mathrm{res}}\;\bm{u}(n-1)),
\end{align}
where $\bm{u}(n) \in \mathbb{C}^{N_n}$ is the internal state vector of RC; $\bm{W}_i \in \mathbb{C}^{N_n \times N_i}$ and $\bm{W}_{\mathrm{res}} \in \mathbb{C}^{N_n \times N_n}$ are the input weights and reservoir weights, respectively; and $f(\cdot)$ is the nonlinear activation function.
The state $\bm{u}(-1)$ is initialized as a zero vector.
The input and reservoir weights are randomly sampled from a uniform distribution and remain unchanged after initialization.
The reservoir weight matrix $\bm{W}_{\mathrm{res}}$ is set to be sparse and have a spectral radius smaller than $1$ to asymptotically eliminate the impact of the initial condition~\cite{jaeger2001echo, tanaka2019recent, lukovsevivcius2012practical}.
The estimated output from RC is obtained by
\begin{align}
    \label{eq:rc_output}
    \hat{\bm{o}}(n) = \bm{W}_{o}\;\tilde{\bm{u}}(n)
\end{align}
where $\hat{\bm{o}}(n) \in \mathbb{C}^{N_o}$ is the estimated output, $N_o$ is the output dimension, $\tilde{\bm{u}}(n) = [\tilde{\bm{y}}(n)^T, \bm{u}(n)^T]^T \in \mathbb{C}^{N_n+N_i}$ is the extended state, and $\bm{W}_{o} \in \mathbb{C}^{N_o\times(N_n+N_i)}$ is the learnable output weight matrix.
After processing the whole sequence, the extended state matrix $\tilde{\bm{U}} \in \mathbb{C}^{(N_n+N_i) \times (L_t+L_f)}$ and estimated output matrix $\hat{\bm{O}} \in \mathbb{C}^{N_o \times (L_t+L_f)}$ can be formed by $\tilde{\bm{U}} \triangleq  [\tilde{\bm{u}}(0), \tilde{\bm{u}}(1), \dots, \tilde{\bm{u}}(L_t+L_f-1)]$ and $\hat{\bm{O}} \triangleq [\hat{\bm{o}}(0), \hat{\bm{o}}(1), \dots, \hat{\bm{o}}(L_t+L_f-1)]$, respectively.

\subsection{Learning algorithm}
Suppose the target output is $\bm{X} \triangleq [\bm{x}(0), \bm{x}(1),\dots,\bm{x}(L_t-1)] \in \mathbb{C}^{N_o \times L_t}$.
The objective function of learning RC is
\begin{align}
    \label{eq:rc_objective1}
    \min_{l_f \in \mathcal{L}_f} \min_{\bm{W}_{o}} ||\hat{\bm{O}}_{l_f}- \bm{X} ||_F^2,
\end{align}
where $\hat{\bm{O}}_{l_f} \triangleq \hat{\bm{O}}[:, l_f:l_f+L_t-1] \in \mathbb{C}^{N_o \times L_t}$ is the truncated estimated output by taking the columns of $\hat{\bm{O}}$ from index $l_f$ to $l_f+L_t-1$, and $l_f$ is a given forget length in the forget length set $\mathcal{L}_f$ with maximum length $L_f$.
By substituting \eqref{eq:rc_output} into \eqref{eq:rc_objective1}, the loss function can be further written as
\begin{align}
    \min_{l_f \in \mathcal{L}_f} \min_{\bm{W}_{o}} ||\bm{W}_{o}\;\tilde{\bm{U}}_{l_f} - \bm{X} ||_F^2,
\end{align}
where $\tilde{\bm{U}}_{l_f} \triangleq \tilde{\bm{U}}[:, l_f:l_f+L_t-1]$ is the truncated extended state matrix.

The objective is learned by alternatively learning the output weights $\bm{W}_{o}$ and the forget length $l_f$.
Specifically, for a given forget length $l_f$, the optimal output weights are acquired by the close-form least square (LS) solution
\begin{align}
    \label{eq:1d_rc_output_mtx_estimation}
    \hat{\bm{W}}_o^{l_f} = \bm{X}\;\tilde{\bm{U}}_{l_f}^\dag.
\end{align}
The optimal forget length is determined by the length that achieves the minimum loss after plugging in the $\hat{\bm{W}}_o^{l_f}$, which can be expressed as
\begin{align}
    \hat{l}_f = \argmin_{l_f \in \mathcal{L}_f} ||\hat{\bm{W}}_o^{l_f}\;\tilde{\bm{U}}_{l_f} - \bm{X} ||_F^2.
\end{align}

\subsection{Testing with 1D-RC}
During the testing stage, the estimated output $\hat{\bm{X}}_{\mathrm{test}} \in \mathbb{C}^{N_o \times L_t}$ is given by
\begin{align}
    \label{eq:1d_rc_test_predict}
    \hat{\bm{X}}_{\mathrm{test}} = \hat{\bm{W}}_o^{\hat{l}_f}\;\tilde{\bm{U}}^{(\mathrm{test})}_{\hat{l}_f},
\end{align}
where $\hat{\bm{W}}_o^{\hat{l}_f}$ is the learned output weight with the optimal forget length $\hat{l}_f$, and $\tilde{\bm{U}}^{(\mathrm{test})}_{\hat{l}_f} = \tilde{\bm{U}}^{(\mathrm{test})}[:, \hat{l}_f:\hat{l}_f+L_t-1]$ is the truncated extended state matrix at the test time using the optimal forget length $\hat{l}_f$.




\section{System Model}
\label{sec:sys_model}

The transmitter and receiver structures in the OTFS system are shown in Fig.~\ref{fig:otfs_system}.
The $Q$ quadrature amplitude modulation ($Q$-QAM) symbols from the modulation alphabet set $\mathcal{A}$ are modulated in the DD domain, which forms the transmitted signal $\boldsymbol{X}$ of size $M\times N$ in the DD domain.
$M$ and $N$ denote the number of delay bins and Doppler bins, respectively.

\begin{figure*}
\centering
\includegraphics[width=0.7\linewidth]{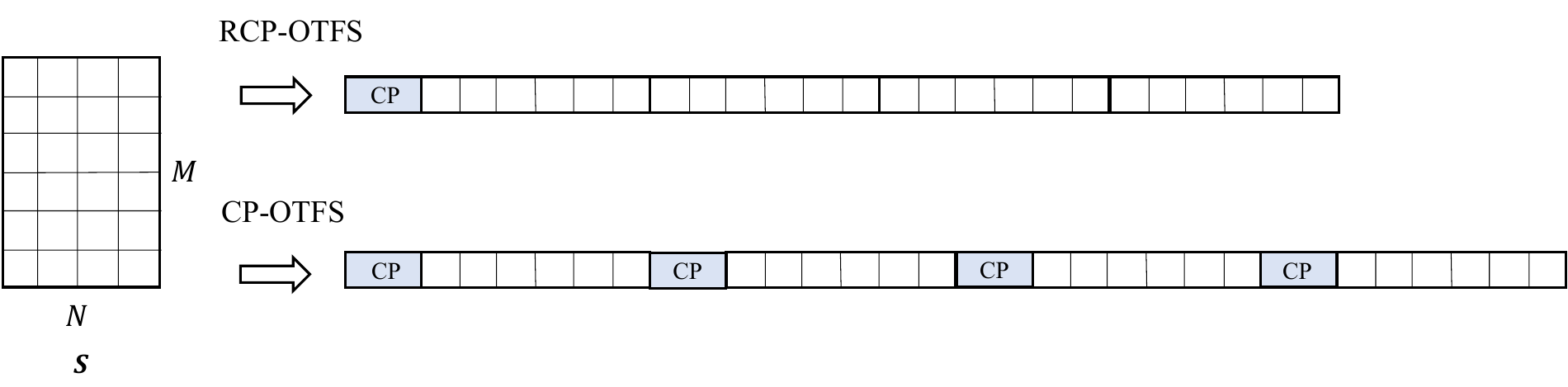}
\caption{OTFS system variants.}
\label{fig:otfs_variants}
\end{figure*}

\subsection{OTFS transmitter and receiver}
The transmitted signal $\boldsymbol{X}$ in the DD domain is converted to the TF domain through inverse symplectic finite Fourier transform (ISFFT) operation, which can be written as
\begin{align}
    \boldsymbol{X}_{tf} = \mathrm{ISFFT}(\boldsymbol{X}) = \boldsymbol{F}_M\boldsymbol{X}\boldsymbol{F}_N^H,
\end{align}
where $\boldsymbol{X}_{tf}$ represent the TF domain signal.
The TF domain signal is then transformed to the time domain signal $\bm{S}\in \mathbb{C}^{M\times N}$ for transmission by the Heisenberg transform.
The transmitted signal can be expressed as
\begin{align}
    \label{eq:dd_to_time}
    \bm{S} = \bm{G}_{tx}\bm{F}_M^H\bm{X}_{tf} = \bm{G}_{tx}\bm{X}\bm{F}_N^H,
\end{align}
where $\bm{G}_{tx} = \mathrm{diag}[g_{tx}(0), g_{tx}(T/M), \dots, g_{tx}((M-1)T/M))] \in \mathbb{C}^{M\times M}$ is a diagonal matrix formed by the samples from the transmit pulse shaping waveform $g_{tx}(t)$ with duration $T$.
When adopting the rectangular pulse shaping, $\bm{G}_{tx}$ is an identity matrix with $\bm{G}_{tx} = \bm{I}_{M}$.
The vector form can be written as $\bm{s} = \mathrm{vec}(\bm{S}) \in \mathbb{C}^{MN\times 1}$.



The received time domain signal $\boldsymbol{r}$ is converted back to the TF domain $\boldsymbol{Y}_{tf}$ through the Winger transform, which can be formulated by
\begin{align}
    \boldsymbol{Y}_{tf} = \bm{F}_M\bm{G}_{rx}\mathrm{vec}^{-1}(\boldsymbol{r}),
\end{align}
where $\bm{G}_{rx} = \mathrm{diag}[g_{rx}(0), g_{rx}(T/M), \dots, g_{rx}((M-1)T/M))] \in \mathbb{C}^{M\times M}$ is formed by the samples from the received pulse-shaping waveform $g_{rx}(t)$.
The DD domain received signal $\boldsymbol{Y}$ is obtained by applying the SFFT to the $\boldsymbol{Y}_{tf}$, which is expressed as
\begin{align}
    \boldsymbol{Y} = \mathrm{SFFT}(\boldsymbol{Y}_{tf}) = \boldsymbol{F}_M^H\boldsymbol{Y}_{tf}\boldsymbol{F}_N.
\end{align}
In this work, we consider the practical rectangular transmit and received pulse shaping waveforms, in which case $\bm{G}_{tx}$ and $\bm{G}_{rx}$ are reduced to the identity matrix, i.e., $\bm{G}_{tx} = \bm{G}_{rx}= \bm{I}_{M}$~\cite{raviteja2018practical}.

\subsection{Channel}
The channel response of the time-varying channel in the DD domain can be represented by
\begin{align*}
    h(\tau, \nu) = \sum_{i=0}^{P-1} h_i \delta(\tau - \tau_i)\delta(\nu - \nu_i),
\end{align*}
where $h_i$, $\tau_i$, and $\nu_i$ represent the complex path gain, delay, and Doppler shift of the $i$-th path; $P$ is the number of propagation paths.
The normalized delay shift $\mathcal{l}_i$ and Doppler shift $\kappa_i$ are given by $\tau_i = \frac{\mathcal{l}_i}{M\Delta f}$ and $\nu_i = \frac{\kappa_i}{NT}$,
where $\mathcal{l}_i$ and $\kappa_i$ are not necessarily integers, and $\Delta f$ is the subcarrier spacing.
In the time domain, the received signal can be expressed as~\cite{hadani2017orthogonal}
\begin{align*}
    r(t) = \int \int h(\tau, \nu)s(t-\tau)e^{j2\pi \nu (t-\tau)}d\tau d\nu + w(t),
\end{align*}
where $s(t)$ denotes the transmitted signal, and $w(t)$ is the additive Gaussian noise.

\subsection{Variants of OTFS system}
\label{sec:variants_otfs_system_io}
We consider two variants of the OTFS system: the RCP-OTFS system and the CP-OTFS system.
As shown in Fig.~\ref{fig:otfs_variants}, in the RCP-OTFS system, a single cyclic prefix (CP) with a length larger than the maximum delay length is added to the beginning of the OTFS subframe to avoid the interference between two consecutive OTFS subframes.
Alternatively, the CP-OTFS system can be implemented as an overlay of the OFDM system, where CP is added for each OFDM symbol in the subframe, i.e., $N$ CPs for one OTFS subframe.
The RCP-OTFS system has a higher spectral efficiency than the CP-OTFS system as only one CP is adopted for the entire subframe~\cite{raviteja2018practical}.
On the other hand, the CP-OTFS system is more compatible with the existing OFDM system, since it can be implemented by adding a pre-processing block and a post-processing block to the OFDM system~\cite{zou2021low, zhou2022learningotfs}.

The input-output relationships in the DD domain of both systems are summarized below.
For ease of discussion, we only show the relationship with integer delay and integer Doppler in~\eqref{eq:rcp_otfs_relationship} and~\eqref{eq:cp_otfs_relationship}.
The relationships with fractional delay and fractional Doppler and the derivation are provided in~\eqref{eq:rcp_fractional} and~\eqref{eq:cp_otfs_fractional} in the Appendix~\ref{appendix:io_relation}.
For simplicity, we omit the noise term.


\subsubsection{RCP-OTFS system}
The input-output relationship in the DD domain for the RCP-OTFS system after adding and removing the CP is given as
\begin{align}
    \label{eq:rcp_otfs_relationship}
   {Y}[l, k] = \sum_{i=0}^{P-1} h_i z^{k_i(\langle l-l_i\rangle_M)}\alpha_{l_i}[l, k] {X}[\langle l-l_i \rangle_M, \langle k-k_i\rangle_N],
\end{align}
where ${Y}[l, k]$ is the $(l, k)$-the element in the received DD-domain signal $\bm{Y}$ with $l=0,1,\dots,M-1$ and $k=0,1,\dots, N-1$; $z$ is defined as $z\triangleq e^{j\frac{2\pi}{NM}}$; $l_i$ and $k_i$ represent the integer delay and integer Doppler; 
the $\alpha_{l_i}[l, k]$ denotes
\begin{align}
    \alpha_{l_i}[l, k] \triangleq \begin{cases}
      e^{-j\frac{2\pi k}{N}}, & \text{if}\ l<l_i \\
      1, & \text{otherwise}.
    \end{cases}
\end{align}

\subsubsection{CP-OTFS system}
The DD-domain input-output relationship in the CP-OTFS system after adding and removing the CP is expressed as
\begin{align}
    \label{eq:cp_otfs_relationship}
    {Y}[l, k] = \sum_{i=0}^{P-1} h_i \tilde{z}^{k_i(N_{cp} + l-l_i)} {X}[\langle l-l_i\rangle_M, \langle k-k_i\rangle_N],
\end{align}
where $\tilde{z} \triangleq e^{j\frac{2\pi}{N(M+N_{cp})}}$ and $N_{cp}$ is the CP length.

As shown in~\eqref{eq:rcp_otfs_relationship} and~\eqref{eq:cp_otfs_relationship}, the difference between the relationship in the RCP-OTFS and the CP-OTFS mainly lies in the phase terms: $z^{k_i(\langle l-l_i\rangle_M)}$ and $\tilde{z}^{k_i(N_{cp} + l-l_i)}$, respectively.
In addition, the relationship in the RCP-OTFS system has an extra phase term $\alpha_{l_i}[l, k]$ that is conditioned on the value of $l$.
In other words, the inter-symbol interference that is not removed in the RCP-OTFS system is lumped into the extra phase term in the DD domain for the detector to handle~\cite{raviteja2018interference}.
While the analysis of phase differences is based on the case with integer delay and integer Doppler, the same observation also applies to the input and output relationship with fractional delay and fractional Doppler, which are shown in the Appendix~\ref{appendix:io_relation}.

From relationships with integer delay and integer Doppler in~\eqref{eq:rcp_otfs_relationship} and~\eqref{eq:cp_otfs_relationship} and relationships with fractional delay and fractional Doppler in~\eqref{eq:rcp_fractional} and~\eqref{eq:cp_otfs_fractional}, we can obtain a general form of the input-output relationship in the DD domain.
Specifically, in general, the input-output relationship in the DD domain of both systems can be written as 
\begin{align}
    \label{eq:otfs_relationship}
   Y[l,k]=\sum_{l'=0}^{M-1}\sum_{k'=0}^{N-1}H_{l, k}[l',k']X[\langle l-l'\rangle_M,\langle k-k'\rangle_N],
\end{align}
where $H_{l, k}[l',k']$ is the effective DD-domain channel.
As shown in \eqref{eq:otfs_relationship}, the channel interaction with the transmitted symbols in the DD domain is a 2D circular operation.

\subsection{Problem formulation}


\begin{figure}
\centering%
\subfloat[]{\label{a}\includegraphics[width=0.4\linewidth]{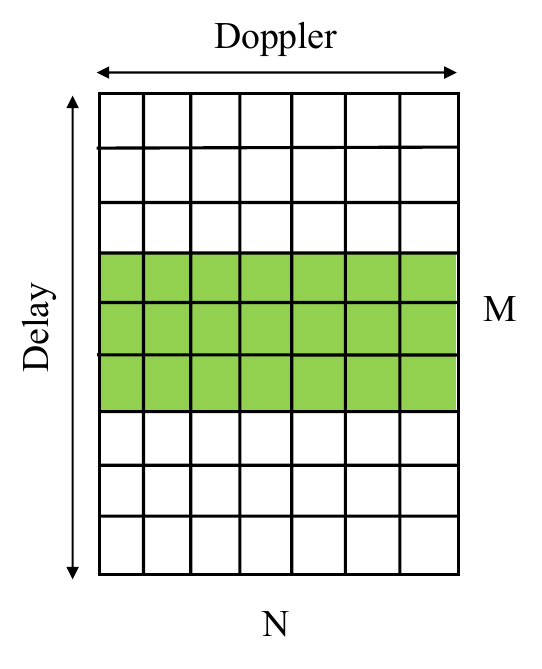}}%
\hspace{1em}
\subfloat[]{\label{b}\includegraphics[width=0.4\linewidth]{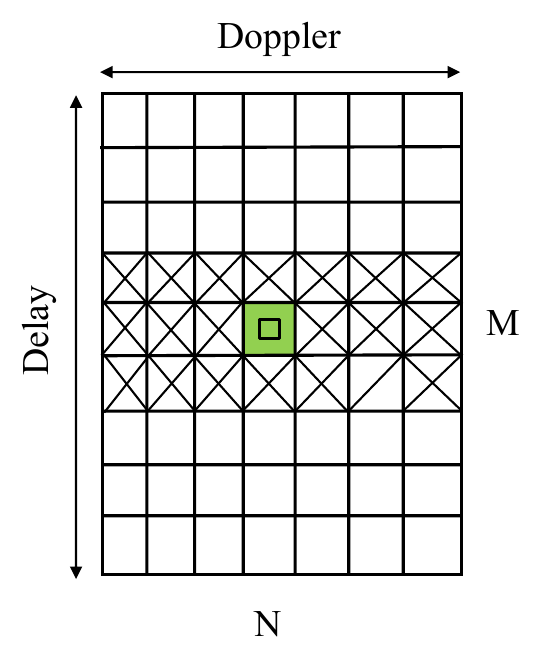}}
\caption{Pilot patterns.
(a) Blockwise pilot pattern.
(b) Spike pilot pattern. 
The green grids are filled with known pilot symbols. 
The green grid with a square marker denotes the spike pilot.
The cross marker represents the guard symbols.
The blank region represents the data symbol position.
\label{figs:pilot_pattern}
}
\end{figure}

\begin{table*}
\centering
\tiny
\caption{Notations appearing in 2D-RC}
\begin{tabular}{ l  l }
\toprule
\textbf{Symbol} & \textbf{Definition} \\
\midrule
$M_w$ & The window size along the delay dimension \\
$N_w$ & The window size along the Doppler dimension \\
$N_i$ & The input size to 2D RC \\
$M_f$ & The maximum forget length along the delay dimension \\
$N_f$ & The maximum forget length along the Doppler dimension \\
$N_n$ & The number of neurons\\
$\mathcal{L}_{m}$ & The delay forget length set with $M_f$ as the maximum delay forget length \\
$\mathcal{L}_{n}$ & The Doppler forget length set with $N_f$ as the maximum Doppler forget length \\
$m_f$ & The forget length in the delay forget length set $\mathcal{L}_{m}$\\
$n_f$ & The forget length in the Doppler forget length set $\mathcal{L}_{n}$ \\
$Y[l, k]$ & The ($l, k$)-th element of the received signal $\boldsymbol{Y}$ \\
$Y_c[l, k]$ & The ($l, k$)-th element of the phase compensated received signal $\boldsymbol{Y}_c$ \\
$\hat{O}[m, n]$ & The estimated ($m, n$)-th output from 2D RC \\
$\tilde{\bm{y}}[m, n]\in \mathbb{C}^{N_i}$ & The ($m, n$)-th element along the second and third dimensions of the input $\tilde{\mathbcal{Y}}$ \\
$\bm{u}[m, n] \in \mathbb{C}^{N_n}$ & The state vector for the $(m, n)$-th input \\
$\tilde{\bm{u}}[m, n] \in \mathbb{C}^{N_n+N_i}$ & The extended state of 2D RC \\
$\bm{Y}_w[l, k] \in \mathbb{C}^{M_w \times N_w}$ & The windowing region for the input $Y[l, k]$ \\
$\bm{Y}_c \in \mathbb{C}^{M \times N}$ & The phase compensated received signal \\
$\bm{W}_i \in \mathbb{C}^{N_n \times N_i}$ & The input weight matrix \\
$\bm{W}_r \in \mathbb{C}^{N_n \times N_n}$ & The reservoir weight matrix along the row axis \\
$\bm{W}_c \in \mathbb{C}^{N_n \times N_n}$ & The reservoir weight matrix along the column axis \\
$\bm{W}_d \in \mathbb{C}^{N_n \times N_n}$ & The reservoir weight matrix along the diagonal axis \\
$\bm{W}_o \in \mathbb{C}^{1\times (N_n+N_i)}$ & The output weight matrix \\
$\hat{\bm{O}} \in \mathbb{C}^{(M + M_f) \times (N + N_f)}$ & The estimated output \\
$\hat{\bm{O}}_{m_f, n_f} \in \mathbb{C}^{M \times N}$ & The truncated output with delay forget length $m_f$ and Doppler forget length $n_f$ \\
$\bar{\bm{U}}_{m_f, n_f} \in \mathbb{C}^{(N_n+N_i) \times MN}$ & The masked truncated extended state matrix formed by vectoring $\bar{\mathbcal{U}}_{m_f, n_f}$ \\
$\tilde{\bm{U}}_{\hat{m}_f, \hat{n}_f} \in \mathbb{C}^{(N_n+N_i) \times MN}$ & The truncated extended state matrix formed by vectoring $\tilde{\mathbcal{U}}_{\hat{m}_f, \hat{n}_f}$ \\
$\hat{\bm{W}}_o^{(m_f, n_f)}$ & The trained output weights when utilizing delay forget length $m_f$ and Doppler forget length $n_f$ \\
$\mathbcal{Y}_w \in \mathbb{C}^{N_i \times M \times N}$ & The 2D windowed input \\
$\tilde{\mathbcal{Y}} \in \mathbb{C}^{N_i \times (M + M_f) \times (N + N_f)}$ & The 2D processed input to the 2D RC \\
${\mathbcal{U}} \in \mathbb{C}^{N_n \times (M + M_f) \times (N + N_f)}$ & The state tensor \\
$\tilde{\mathbcal{U}} \in \mathbb{C}^{(N_n+N_i) \times (M + M_f) \times (N + N_f)}$ & The extended state tensor \\
$\tilde{\mathbcal{U}}_{m_f, n_f} \in \mathbb{C}^{(N_n+N_i) \times M \times N}$ & The truncated extended state tensor \\
$\bar{\mathbcal{U}}_{m_f, n_f} \in \mathbb{C}^{(N_n+N_i) \times M \times N}$ & The masked truncated extended state tensor \\

\bottomrule
\end{tabular}
\label{tab:notations}
\end{table*}

The symbol detection task in the OTFS system is to recover the transmitted DD-domain symbol $\bm{X}$ in one OTFS subframe from the received signal $\bm{r}$.
In this work, we consider a practical setting, where the perfect CSI is not available. 

To aid the detection of the unknown data symbols, the pilot symbols, which are known at both the transmitter and receiver sides, are inserted in each subframe.
In this paper, we consider two pilot structures for symbol detection approaches in the OTFS system: the blockwise pilot pattern and the spike pilot pattern, which are shown in Fig.~\ref{figs:pilot_pattern}.
For learning-based approaches in the OTFS system, the blockwise pilot structure is adopted in the delay-Doppler domain, where pilot symbols are placed in a block of the subframe.
For model-based schemes in the OTFS system that require knowledge of the CSI, the spike pilot pattern is utilized for channel estimation~\cite{raviteja2019embedded}.
Specifically, a spike pilot is transmitted along with guard symbols surrounding it.
The guard symbols are set to occupy the full Doppler axis following the pilot pattern introduced in~\cite{raviteja2019embedded}.
The spike pilot is placed in the middle of the pilot region, as shown in Fig.~\ref{figs:pilot_pattern} (b).
The pilot occupancy for the spike pilot pattern includes both the spike pilot and guard symbols.
All the considered pilot patterns are set to have the same pilot overhead.

Different pilot structures are adopted for the learning-based approaches and model-based methods in the OTFS system.
The reason is that model-based approaches need to avoid interference between pilot symbols and data symbols to have an accurate channel estimation.
In contrast, learning-based approaches need to learn from cases when such interference is present to prevent model mismatches during training and testing.
Furthermore, to avoid discrepancies between pilot and data symbols that could lead to training and testing model mismatches, pilot symbols are sampled randomly from the modulation alphabet set rather than being set as a spike.
Therefore, the superimposed pilot pattern in~\cite{yuan2021data}, which overlays a spike pilot onto data symbols, is not considered for learning-based methods in this paper.
More details about the choice of pilot patterns for learning-based and model-based approaches are provided in~\cite{zhou2022learningotfs, xu2023DetectToLearn}.
It is noteworthy that other alternative interleaved and superimposed pilot patterns have already been investigated in~\cite{zhou2022learningotfs}, which do not show comparable performance to the blockwise pilot pattern for the 1D-RC method.
For a fair comparison with the 1D-RC approach and paper conciseness, we mainly focus on the blockwise pilot pattern.
Denote $\bm{\Omega}$ as the pilot position indication matrix with $1$ indicating the pilot positions and $0$ specifying the data position.
For the introduced learning-based approach, the input to the NN is the received DD-domain signal $\bm{Y}$, which is obtained by transforming the time-domain signal $\bm{r}$ into the DD domain.
The training target is composed of the pilot symbols modulated in the DD domain.
Therefore, the training dataset within one subframe can be written as
\begin{align*}
\{ \bm{Y}, \bm{X}_{\mathrm{train}} \triangleq \bm{\Omega} \odot \bm{X}\}.
\end{align*}
Accordingly, the testing dataset can be obtained by 
\begin{align*}
\{ \bm{Y}, \bm{X}_{\mathrm{test}} \triangleq \bar{\bm{\Omega}} \odot \bm{X}\},
\end{align*}
where $\bar{\bm{\Omega}}$ is the complement of $\bm{\Omega}$.


\section{Introduced Approach}
\label{sec:introduced_approach}
In this section, we introduce the 2D-RC approach for online subframe-based symbol detection in the OTFS system.
The introduced 2D-RC retains the same simple training process as RC, enabling it to perform online symbol detection with limited training pilots on a subframe basis.
Moreover, it is uniquely designed to facilitate online symbol detection tailored towards the OTFS system.
Specifically, the DD-domain channel works as a 2D circular operation over transmitted symbols in the OTFS system as shown in~\eqref{eq:otfs_relationship}.
To equalize this 2D circular channel effect, 2D-RC is designed to have a 2D circular padding procedure and a 2D filtering structure.
By embedding the domain knowledge of the OTFS system, 2D-RC can work in the DD domain with only a single NN for detection, as opposed to the 1D-RC approach~\cite{zhou2022learningotfs} that exploits multiple RCs to track the channel variations in the time domain.
It is noteworthy that the incorporated 2D circular operation exists in the DD-domain input-output relationship in general regardless of the exact channel model.
Therefore, the specific channel model does not change the design of the 2D-RC algorithm.
Notations are summarized in Tab.~\ref{tab:notations}.


\subsection{Pre-processing}
The introduced 2D-RC conducts the detection process in the DD domain.
Therefore, the input is the received signal $\boldsymbol{Y} \in \mathbb{C}^{M\times N}$ in the DD domain.
Similar to 1D-RC, the pre-processing procedures, including windowing and padding, are also adopted before the processing of 2D-RC.
The difference is that the pre-processing steps for 2D-RC are conducted in a 2D way.
Furthermore, based on the input-output relationship, we add the phase compensation step for the RCP-OTFS system.

\subsubsection{Phase Compensation}

As shown in \eqref{eq:rcp_otfs_relationship}, the input-output relationship in the RCP-OTFS system has an extra phase term that is conditioned on the delay index of the received signal.
The extra phase term may result in a training and testing mismatch when adopting the block pilot pattern.
The phase change may not be captured during the training stage when the block pilots are placed in the middle of the OTFS subframe.
Therefore, for the RCP-OTFS system, we add a phase compensation step to roughly compensate for the phase change in the received signal.
Specifically, the received signal after phase compensation can be written as
\begin{align}
      Y_c[l, k] \triangleq \begin{cases}
      Y[l, k]e^{j\frac{2\pi k}{N}}, & \text{if}\ l<l_c \\
      Y[l, k], & \text{otherwise},
    \end{cases}
\end{align}
where $Y_c[l, k]$ and $Y[l, k]$ are the $(l, k)$-th element in the phase-compensated received signal $\bm{Y}_c$ and the received signal $\boldsymbol{Y}$, respectively; $l = 0, 1, \dots, M-1$ and $k = 0, 1, \dots, N-1$; and $l_c$ is a tunable parameter.
For the CP-OTFS system, the phase compensation step is skipped and we have $\bm{Y}_c = \bm{Y}$.

\begin{figure*}
\centering
\includegraphics[width=0.7\linewidth]{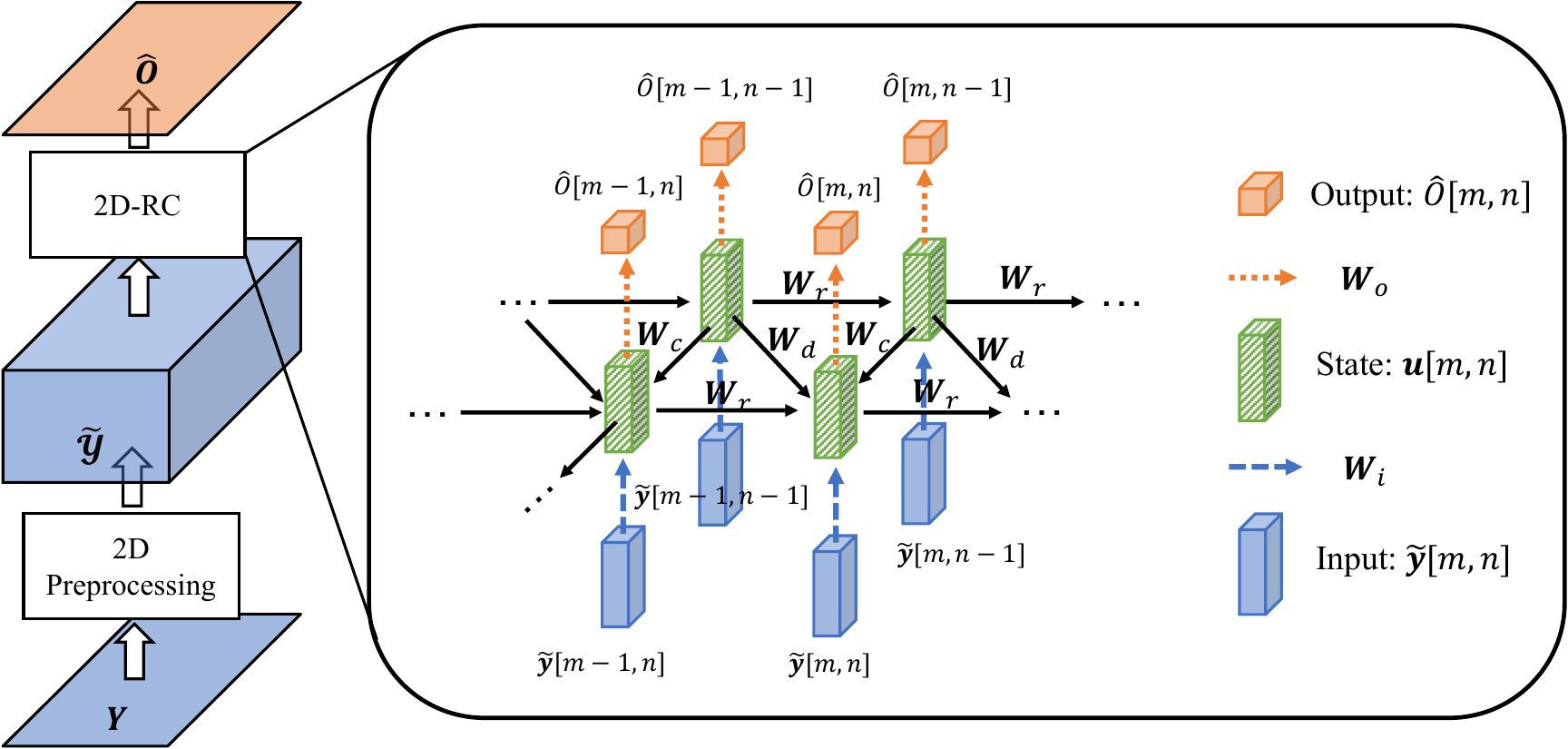}
\caption{2D-RC Structure. For simplicity, the nonlinear function and the extended state are ignored here. 
}
\label{figs:2d_rc_structure}
\end{figure*}

\begin{figure}
\centering
\includegraphics[width=0.7\linewidth]{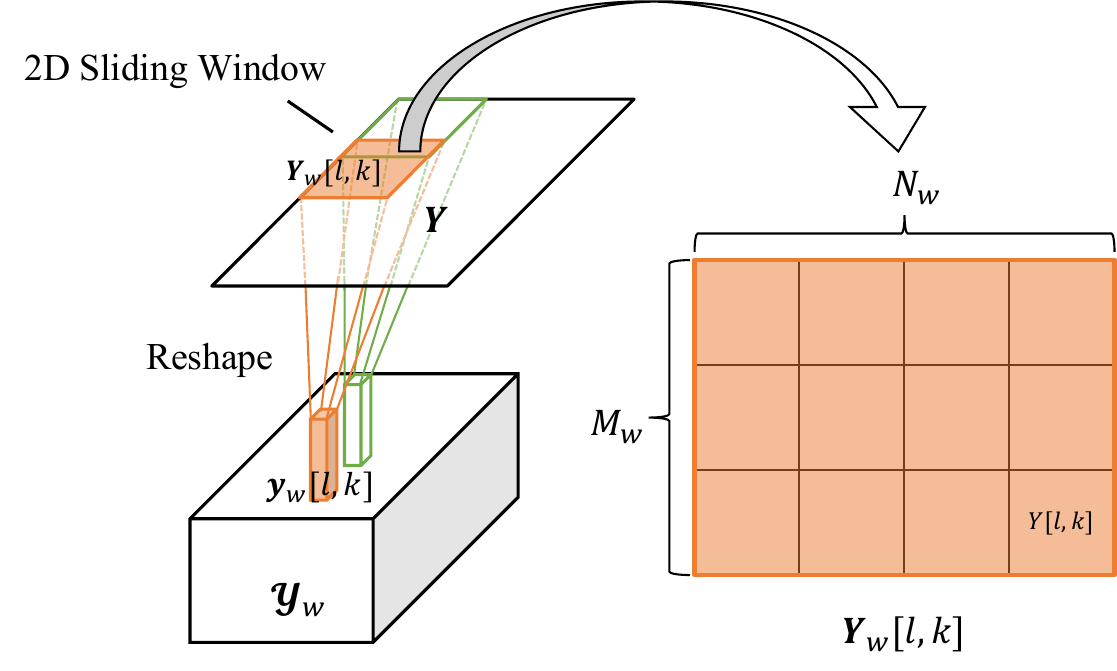}
\caption{The windowing process in 2D-RC.}
\label{figs:2d_rc_pre_process}
\end{figure}

\subsubsection{2D Windowing}
We adopt a 2D sliding window with size $M_w \times N_w$ to process the input, where $M_w$ is the window size along the delay dimension and $N_w$ is the window size along the Doppler dimension.
For each ${Y}_c[l, k]$, the windowing region is obtained by $\bm{Y}_w[l, k] = \boldsymbol{Y}_c[l-M_w+1:l, k-N_w+1:k] \in \mathbb{C}^{M_w \times N_w}$.
When the $l<M_w-1$ or $k<N_w-1$, zeros are filled in the windowing region to maintain the window size of $M_w \times N_w$.
The windowed input is formed by $\bm{y}_w[l, k] = \mathrm{vec}(\mathrm{rev}(\bm{Y}_w[l, k]^T)) \in \mathbb{C}^{N_i}$, where $\mathrm{rev}(\cdot)$ stands for reserving the values in the matrix along both dimensions, $\mathrm{vec}(\cdot)$ represents vectoring the matrix by stacking along the columns, and $N_i = M_wN_w$.
By collecting all the $\bm{y}_w[l, k]$, we obtain an input tensor $\mathbcal{Y}_w \in \mathbb{C}^{N_i \times M \times N}$.
Fig.~\ref{figs:2d_rc_pre_process} visualizes the 2D windowing process.


\subsubsection{2D Circular Padding}
As in 1D-RC, 2D-RC also needs to learn the optimal forget length to eliminate the impact of the initial state.
Based on the padding process in 1D-RC, we design a 2D circular padding process to facilitate the learning process of the optimal forget length.
Let $M_f$ and $N_f$ be the maximum forget length along the delay and Doppler dimension, respectively.
The 2D padded input $\tilde{\mathbcal{Y}} \in \mathbb{C}^{N_i \times (M + M_f) \times (N + N_f)}$ is obtained by concatenating the $\mathbcal{Y}_w$ along the second and third dimensions as follows:
\begin{align*}
    \tilde{\mathbcal{Y}} = & \nonumber
    \text{cat}_2( \text{cat}_3(\mathbcal{Y}_w,  \;\;           \mathbcal{Y}_w[:, :, 0:N_f-1]),\nonumber\\
               & \text{cat}_3(\mathbcal{Y}_w[:, 0:M_f-1, :],\mathbcal{Y}_w[:, 0:M_f-1, 0:N_f-1] )).
\end{align*}
Note that this padding process is different from the zero padding process for 1D-RC in Sec.~\ref{sec:prelim_pre_process}.
The circular padding is employed in 2D-RC, where the values at the start are utilized to pad at the end of the corresponding dimension.
The reason is that in the OTFS system, the received signal is acquired through a 2D circular operation between the channel and the input signal in the DD domain.
The circular operation in the input-output relationship inspires the utilization of the 2D circular padding.




\subsection{Structure of 2D-RC}

Denote $\tilde{\bm{y}}[m, n]\in \mathbb{C}^{N_i}$ as the ($m, n$)-th element along the second and third dimensions of the pre-processed input $\tilde{\mathbcal{Y}}$, where $m = 0, 1, \dots, M+M_f-1$ and $n = 0, 1, \dots, N+N_f-1$.
We design the state transition equation for 2D-RC as
\begin{align}
    \label{eq:2d_rc_state_transition}
    \bm{u}[m, n] &= f(\bm{W}_i\;\tilde{\bm{y}}[m, n] + \bm{W}_r\;\bm{u}[m-1, n] \nonumber\\
    &+ \bm{W}_d\;\bm{u}[m-1, n-1] + \bm{W}_c\;\bm{u}[m, n-1]),
\end{align}
where $\bm{u}[m, n] \in \mathbb{C}^{N_n}$ represent state vector for the $(m, n)$-th input; $N_n$ stands for the number of neurons; $\bm{W}_i \in \mathbb{C}^{N_n \times N_i}$ is the input weight matrix; $N_i$ denote the input dimension; $\bm{W}_r \in \mathbb{C}^{N_n \times N_n}$, $\bm{W}_c \in \mathbb{C}^{N_n \times N_n}$, and $\bm{W}_d \in \mathbb{C}^{N_n \times N_n}$ denote the reservoir weights along the row, column, and diagonal directions, respectively; $f(\cdot)$ is the nonlinear activation function.
The input weights and reservoir weights are all randomly initialized by sampling from a uniform distribution.
In line with the 1D-RC approach, all reservoir weights are configured to be sparse with spectral radii less than $1$. 
The initial states $\bm{u}[-1, n]$, $\bm{u}[m, -1]$, and $\bm{u}[-1, -1]$ are all initialized as zero vectors.
The output equation is formulated as
\begin{align}
    \hat{O}[m, n] = \bm{W}_o\;\tilde{\bm{u}}[m, n],\label{eq:2d_rc_output}\\
    \tilde{\bm{u}}[m, n] = \begin{bmatrix}
           \tilde{\bm{y}}[m, n] \\
            \bm{u}[m, n]
         \end{bmatrix},
\end{align}
where $\tilde{\bm{u}}[m, n]\in \mathbb{C}^{N_n+N_i}$ is the extended state formed by concatenating the input and the state, $\bm{W}_o \in \mathbb{C}^{1\times (N_n+N_i)}$ stands for the output weights.
By collecting all the state vectors ${\bm{u}}[m, n]$, the extended state vectors $\tilde{\bm{u}}[m, n]$, and the estimated output $\hat{O}[m, n]$, we can obtain the state tensor ${\mathbcal{U}} \in \mathbb{C}^{N_n \times (M + M_f) \times (N + N_f)}$, the extended state tensor $\tilde{\mathbcal{U}} \in \mathbb{C}^{(N_n+N_i) \times (M + M_f) \times (N + N_f)}$ and the estimated output matrix $\hat{\bm{O}} \in \mathbb{C}^{(M + M_f) \times (N + N_f)}$.
The structure is shown in Fig.~\ref{figs:2d_rc_structure}.


\subsection{Learning Algorithm}
Like 1D-RC, only the output weights are learned during training.
The training loss for 2D-RC is given as
\begin{align}
\min_{m_f \in \mathcal{L}_{m}, n_f \in \mathcal{L}_{n}} \min_{\bm{W}_{o}} ||\bm{\Omega} \odot \hat{\bm{O}}_{m_f, n_f} - \bm{X}_{\mathrm{train}} ||_F^2, 
\end{align}
where $\hat{\bm{O}}_{m_f, n_f} = \hat{\bm{O}}[m_f:m_f+M-1, n_f:n_f+N-1] \in \mathbb{C}^{M \times N}$ represents the truncated output, $m_f$ is a forget length in the delay forget length set $\mathcal{L}_{m}$ with $M_f$ as the maximum delay forget length, and $n_f$ is a forget length in the Doppler forget length set $\mathcal{L}_{n}$ with $N_f$ as the maximum Doppler forget length.
By vectorizing the output and the target, the training objective can be further written as
\begin{align}
\min_{m_f \in \mathcal{L}_{m}, n_f \in \mathcal{L}_{n}} \!\!\! \min_{\bm{W}_{o}} ||\mathrm{vec}(\bm{\Omega} \odot \hat{\bm{O}}_{m_f, n_f}) - \mathrm{vec}(\bm{X}_{\mathrm{train}}) ||_2^2.
\label{eq:2d_objective_v1}
\end{align}
Let $\tilde{\mathbcal{U}}_{m_f, n_f} = \tilde{\mathbcal{U}}[:, m_f:m_f+M-1, n_f:n_f+N-1] \in \mathbb{C}^{(N_n+N_i) \times M \times N}$ be the truncated extended state.
The masked truncated extended state tensor is denoted as $\bar{\mathbcal{U}}_{m_f, n_f} = \bm{\Omega} \odot_{2} \tilde{\mathbcal{U}}_{m_f, n_f}$, where $\odot_{2}$ represents conducting the Hadamard product along the second and third dimensions.
The masked truncated extended state matrix $\bar{\bm{U}}_{m_f, n_f} = \mathrm{vec}_2(\bar{\mathbcal{U}}_{m_f, n_f}) \in \mathbb{C}^{(N_n+N_i) \times MN}$ is formed by vectoring the last two dimensions of $\bar{\mathbcal{U}}_{m_f, n_f}$ with $\mathrm{vec}_2(\cdot)$ denoting vectoring along the second and third dimensions.
Then by substituting \eqref{eq:2d_rc_output} into \eqref{eq:2d_objective_v1}, the objective function becomes
\begin{align}
    \min_{m_f \in \mathcal{L}_{m}, n_f \in \mathcal{L}_{n}} \min_{\bm{W}_{o}} ||\bm{W}_o\bar{\bm{U}}_{m_f, n_f} - (\mathrm{vec}(\bm{X}_{\mathrm{train}}))^T||_2^2.
\end{align}

Following the training strategy in 1D-RC, the forget length and the output weights are learned alternatively.
We first fix the forget length $m_f$ and $n_f$ and obtain the trained output weights by the LS solution
\begin{align}
    \label{eq:2d_output_mtx_estimation}
    \hat{\bm{W}}_o^{(m_f, n_f)} =  (\mathrm{vec}(\bm{X}_{\mathrm{train}}))^T\bar{\bm{U}}_{m_f, n_f}^\dag.
\end{align}
Then the optimal forget lengths along the delay dimension and Doppler dimension are learned by finding the length that minimizes the loss after plugging in the $\hat{\bm{W}}_o^{(m_f, n_f)}$, i.e.,
\begin{align}
    \hat{m}_f, \hat{n}_f \!= \!\!\!\!\argmin_{m_f \in \mathcal{L}_{m}, n_f \in \mathcal{L}_{n}} \!\!||\hat{\bm{W}}_o^{(m_f, n_f)} \bar{\bm{U}}_{m_f, n_f} \!- \!(\mathrm{vec}(\bm{X}_{\mathrm{train}}))^T||_2^2.
\end{align}
Instead of searching through all the possible delay and Doppler forget length pairs, we first find the optimal Doppler forget length and then find the optimal delay forget length to reduce the training complexity.

\subsection{Testing with 2D-RC}
At the testing stage, the transmitted symbols $\hat{\bm{x}} \in \mathbb{C}^{1 \times MN}$ are estimated by
\begin{align}
    \label{eq:2d_rc_output_eq}
    \hat{\bm{x}} = \mathcal{Q}(\hat{\bm{W}}_o^{(\hat{m}_f, \hat{n}_f)} \;\tilde{\bm{U}}_{\hat{m}_f, \hat{n}_f}),
\end{align}
where $\hat{\bm{W}}_o^{(\hat{m}_f, \hat{n}_f)}$ is the trained output matrix when utilizing the forget length $\hat{m}_f$ and $\hat{n}_f$, $\tilde{\bm{U}}_{\hat{m}_f, \hat{n}_f} =\mathrm{vec}_2(\tilde{\mathbcal{U}}_{\hat{m}_f, \hat{n}_f})  \in \mathbb{C}^{(N_n+N_i) \times MN}$ is obtained by vectoring the truncated extended state tensor $\tilde{\mathbcal{U}}_{\hat{m}_f, \hat{n}_f}$ with forget length $\hat{m}_f$ and $\hat{n}_f$, and $\mathcal{Q}(\cdot)$ is the quantization operation that maps the output to the constellation points.
The transmitted data symbols are extracted with
\begin{align}
    \hat{\bm{X}}_{\mathrm{data}} = \bar{\bm{\Omega}} \odot \hat{\bm{X}},
\end{align}
where $\hat{\bm{X}} =\mathrm{vec}^{-1}(\hat{\bm{x}}) \in \mathbb{C}^{M \times N}$ is the matrix formed by filling the matrix column by column.




\section{Complexity Analysis}
\label{sec:complexity}

\begin{table*}[!t]
\centering
\scriptsize
\caption{Computation Complexity}
\begin{tabular}{ c  c  c }
\toprule
Method & Training\textbackslash{}Channel Estimation & Testing\textbackslash{}Detection  \\
\midrule

LMMSE & $\mathcal{O}(\eta MN)$ & $\mathcal{O}(M^3N^3)$ \\
Low-complexity LMMSE & $\mathcal{O}(\eta MN)$ & $\mathcal{O}(MN\tilde{P}logN)$ \\
MPA & $\mathcal{O}(\eta MN)$ & $\mathcal{O}(N_{\mathrm{iter}}|\mathcal{A}|\tilde{P}MN)$ \\
LSMR-based approach & $\mathcal{O}(\eta MN)$ & $\mathcal{O}(IK\tilde{P}MN)$ \\
1D-RC ($N_i+N_n \leq \eta MN/V$) & $\mathcal{O}(N_n(N_i+N_n)MN +(N_i+N_n)((\eta MN)^2/V+\eta MN)|\mathcal{L}_f|)$ & $\mathcal{O}((N_n+N_i)MN)$   \\
1D-RC ($N_i+N_n > \eta MN/V$) & $\mathcal{O}(N_n(N_i+N_n)MN + (N_i+N_n)((N_i+N_n)\eta MNV+\eta MN)|\mathcal{L}_f|)$ & $\mathcal{O}((N_n+N_i)MN)$   \\
2D-RC & $\mathcal{O}(N_n(N_i+3N_n)MN + (N_i+N_n)((\eta MN)^2+\eta MN)(|\mathcal{L}_{m}|+|\mathcal{L}_{n}|))$ & $\mathcal{O}((N_n+N_i)MN)$ \\

\bottomrule
\end{tabular}
\label{tab:complexity_table}
\end{table*}


In this section, we analyze the computational complexity of 2D-RC and compare it with existing approaches for the OTFS system.
We focus on the computational complexity of matrix multiplication and pseudo-inverse.
The computational cost for matrix addition is ignored here as they are negligible compared to matrix multiplication and inverse.
Note that the complexity for the pseudo-inverse of a matrix with size $M\times N$ ($M<N$) is $\mathcal{O}(MN^2)$ when implemented with the singular value decomposition.
For ease of discussion, we denote the pilot overhead as $\eta = \frac{|\bm{\Omega}|}{MN}$, where $|\bm{\Omega}|$ denotes the number of ones within the pilot mask $\bm{\Omega}$.
Note that guard symbols in the spike pilot pattern are considered as pilot positions and therefore counted in the pilot overhead.

The training complexity of RC consists of two parts: the state transition and the output weights estimation.
The state transition in \eqref{eq:2d_rc_state_transition} has a total complexity of $\mathcal{O}(N_n(N_i+3N_n)(M+M_f)(N+N_f))\approx \mathcal{O}(N_n(N_i+3N_n)MN)$.
The output matrix estimation is obtained by computing the pseudo-inverse of the extended state followed by the multiplication of the target and the inverse of the extended state, as shown in \eqref{eq:2d_output_mtx_estimation}.
As $N_i+N_n < \eta MN$ in practice, the complexity for calculating the pseudo-inverse of the extended state in \eqref{eq:2d_output_mtx_estimation} is $\mathcal{O}((N_n+N_i)(\eta MN)^2)$ for the given forget lengths along delay and Doppler dimensions.
The computational complexity for the matrix multiplication in \eqref{eq:2d_output_mtx_estimation} is $\mathcal{O}(\eta MN(N_i+N_n))$.
Therefore, the output weights estimation process in \eqref{eq:2d_output_mtx_estimation} has a complexity of $\mathcal{O}((N_i+N_n)((\eta MN)^2+\eta MN))$.
When considering the forget length searching process, the complexity becomes $\mathcal{O}((N_i+N_n)((\eta MN)^2+\eta MN)(|\mathcal{L}_{m}|+|\mathcal{L}_{n}|))$, where the $|\mathcal{L}_{m}|$ and $|\mathcal{L}_{n}|$ denote the cardinality of the set $\mathcal{L}_{m}$ and $\mathcal{L}_{n}$, respectively.
The total training complexity is $\mathcal{O}(N_n(N_i+3N_n)MN + (N_i+N_n)((\eta MN)^2+\eta MN)(|\mathcal{L}_{m}|+|\mathcal{L}_{n}|))$.
During the testing stage, only the output estimation step in \eqref{eq:2d_rc_output_eq} needs to be considered, as the states are pre-computed at the training stage.
Therefore, the total testing complexity of the 2D-RC is $\mathcal{O}((N_i+N_n)MN)$.


For the 1D-RC approach in~\cite{zhou2022learningotfs}, multiple 1D-RCs are adopted for detection, where each RC is utilized to learn a local channel feature.
When considering the windowing and padding, the state transition processes for $V$ number of 1D-RCs have a total complexity of $\mathcal{O}(N_n(N_i+N_n)(MN/V+L_f)V)\approx \mathcal{O}(N_n(N_i+N_n)MN)$.
For the output matrix estimation process of each 1D-RC, we consider two cases: (1) $N_i+N_n \leq \eta MN/V$; (2) $N_i+N_n > \eta MN/V$.
When $N_i+N_n \leq \eta MN/V$, the matrix pseudo inverse in \eqref{eq:1d_rc_output_mtx_estimation} has a complexity of $\mathcal{O}((N_i+N_n)(\eta MN)^2/V^2)$.
The complexity of the matrix multiplication in \eqref{eq:1d_rc_output_mtx_estimation} is $\mathcal{O}((N_i+N_n)\eta MN/V)$.
Then the total computational complexity of the output matrix estimation in \eqref{eq:1d_rc_output_mtx_estimation} is $\mathcal{O}((N_i+N_n)((\eta MN)^2/V^2+\eta MN/V))$.
When considering the forget length learning process and $V$ number of 1D-RCs, the complexity becomes $\mathcal{O}(|\mathcal{L}_f|(N_i+N_n)((\eta MN)^2/V+\eta MN))$, where $|\mathcal{L}_f|$ is the number of forget length in the set $\mathcal{L}_f$.
Thus, the total training complexity is $\mathcal{O}((N_i+N_n)(N_nMN +|\mathcal{L}_f|(\eta MN)^2/V+|\mathcal{L}_f|\eta MN))$ in the case of $N_i+N_n \leq \eta MN/V$.
Similarly, when $V$ is large enough to have $N_i+N_n > \eta MN/V$, i.e., a large number of 1D-RCs is adopted, we can obtain the total training complexity as $\mathcal{O}((N_i+N_n)(N_nMN +|\mathcal{L}_f|(N_i+N_n)\eta MNV+|\mathcal{L}_f|\eta MN))$, which is proportional to the number of RCs.
The total testing complexity is $\mathcal{O}((N_n+N_i)MN)$, as the internal states of RC are all pre-computed at the training stage and only the output estimation process is conducted.

The MPA~\cite{raviteja2018interference}, LSMR-based approach~\cite{qu2021TCOMLowcomplexSD}, and LMMSE detector require channel knowledge for detection.
As discussed in~\cite{zhou2022learningotfs}, the complexity of channel estimation with the approach in~\cite{raviteja2019embedded} is $\mathcal{O}(\eta MN)$.
The testing complexity of MPA is $\mathcal{O}(N_{iter}|\mathcal{A}|\tilde{P}MN)$, where $N_{iter}$ is the number of iterations, $|\mathcal{A}|$ is the size of the modulation alphabet set, and $\tilde{P}$ is the total number of estimated taps including the number of estimated virtual taps due to the effect of fractional delay and fractional Doppler.
The LSMR-based method in~\cite{qu2021TCOMLowcomplexSD} has a testing complexity of $\mathcal{O}(IK\tilde{P}MN)$, where $I$ denotes the number of iterations for LSMR and $K$ is the number of iterations for interference cancellation.
The direct implementation of the LMMSE approach has a computational complexity in the order of $\mathcal{O}(M^3N^3)$~\cite{tiwari2019low}.
In~\cite{zou2021low}, the complexity of the LMMSE detector can be reduced to $\mathcal{O}(MN\tilde{P}logN)$.

The computational complexities of different detection schemes are summarized in Tab.~\ref{tab:complexity_table}.
While 2D-RC and 1D-RC may be set with different parameters depending on the simulation performance, the training and testing complexities of these two approaches are in the same order of magnitude.
Furthermore, when $N_i+N_n < \tilde{P}N_{iter}|\mathcal{A}|$, $N_i+N_n < \tilde{P}IK$, and $N_i+N_n < \tilde{P}logN$, RC-based approaches can have lower testing computational costs than MPA, LSMR-based approach, and low-complexity LMMSE, respectively.


\section{Numerical Experiments}
\label{sec:experiments}

\begin{figure}
\centering
\includegraphics[width=0.8\linewidth]{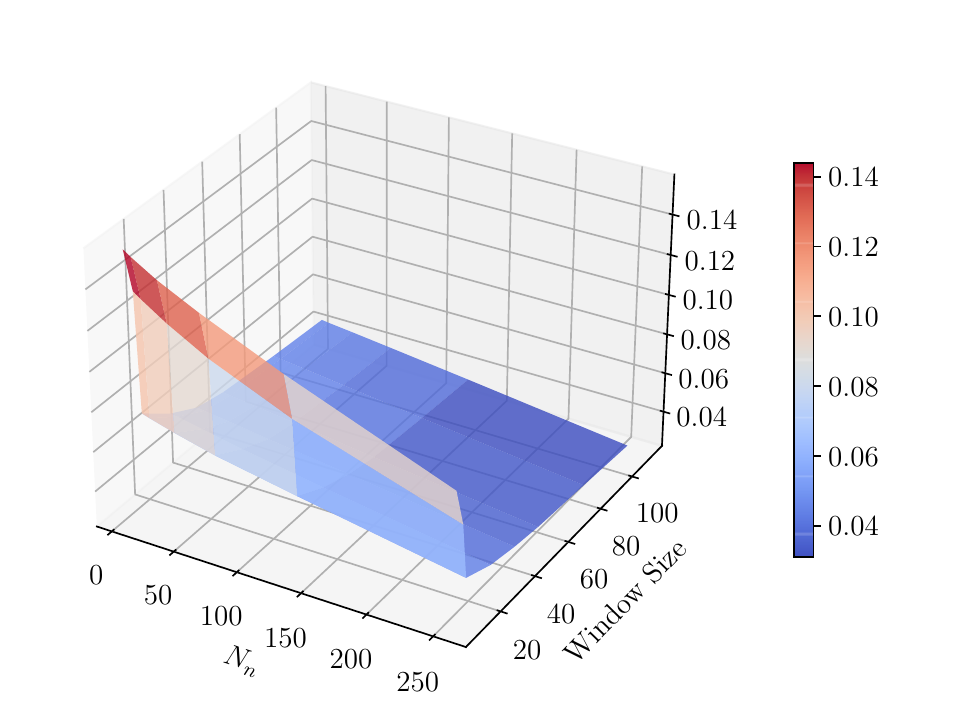}
\caption{Training NMSE with different numbers of neurons and window sizes.}
\label{figs:train_nmse}
\end{figure}



In this section, we evaluate the performance of 2D-RC for symbol detection in the OTFS system.
Unless otherwise specified, we consider the uncoded OTFS system.
We adopt $N=14$ following the 3GPP 5G NR standard~\cite{std3gpp38211}.
The number of subcarriers is set as $M=1024$.
The carrier frequency is $4$ GHz and subcarrier spacing is $15$ KHz.
The 3GPP 5G NR clustered delay line (CDL) channel with delay profile ``CDL-C"~\cite{std3gpp38901} is considered.
The delay spread is $10$ ns.
Unless otherwise specified, the user velocity is set as $150$ km/h.
The pilot overhead is $4.69 \%$, which is set to satisfy the pilot overhead requirement specified in~\cite{std3gpp38211, std3gpp38212}.
With the pilot overhead, the number of delay grids occupied by pilot symbols for both the blockwise pilot pattern and the spike pilot pattern is 48.
All the compared approaches adopt the same training overhead for a fair comparison.



\begin{figure}
\centering
\includegraphics[width=0.8\linewidth]{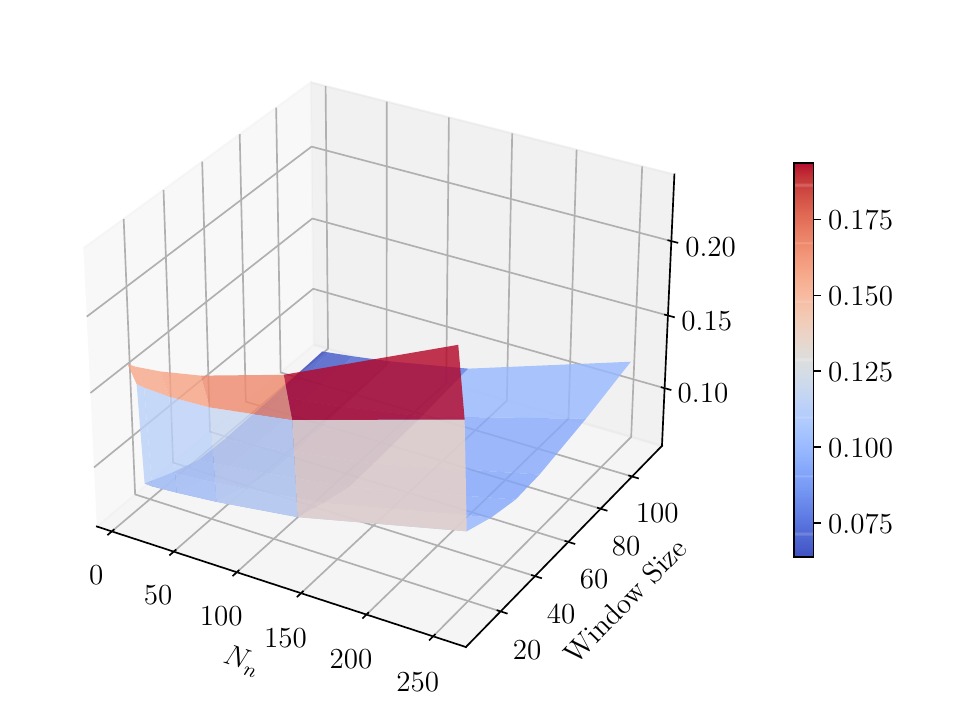}
\caption{Testing NMSE with different numbers of neurons and window sizes.}
\label{figs:test_nmse}
\end{figure}

In Fig.~\ref{figs:train_nmse} and Fig.~\ref{figs:test_nmse}, we investigate how the number of neurons and the window size affect the training normalized mean square error (NMSE) and testing NMSE of 2D-RC.
As shown in Fig.~\ref{figs:train_nmse}, the training NMSE exhibits a decreasing trend with the increase of both the number of neurons and the window size. 
This observation can be attributed to the fact that as the number of neurons increases, the 2D-RC model is capable of mapping the input to a higher-dimensional state space, consequently expanding the model capacity.
Furthermore, the windowing operation employed on the input can be interpreted as incorporating multiple skip connections within the neural network architecture, as discussed in~\cite{zhou2022learningotfs}.
The presence of skip connections behaves as multiple ensembles of NN models, further increasing the model capacity~\cite{veit2016residual}.
The increased model capacity enables the 2D-RC to capture more complex patterns from the input data, leading to a lower training NMSE.
However, due to overfitting, the testing NMSE increases when the model capacity is too large, as shown in Fig.~\ref{figs:test_nmse}.
Therefore, there is a trade-off between the number of neurons and the window size.
Based on the above analysis and the simulation, the parameters of 2D-RC are set as $N_n=6$, $M_w=4$, $N_w=14$, and $l_c=7$.
The delay forget length and Doppler forget length are searched in the range of $7$ to $8$ and the range of $13$ to $14$, respectively.
The spectral radii of all the reservoir weights are configured as $0.9$ and the sparsities are set as $0.6$.
The parameters of 2D-RC are empirically determined through simulations.
The nonlinear activation function is selected as the hyperbolic tangent function.
The quantization operation is set as the nearest neighbor mapping.


\begin{figure}
\centering
\includegraphics[width=0.8\linewidth]{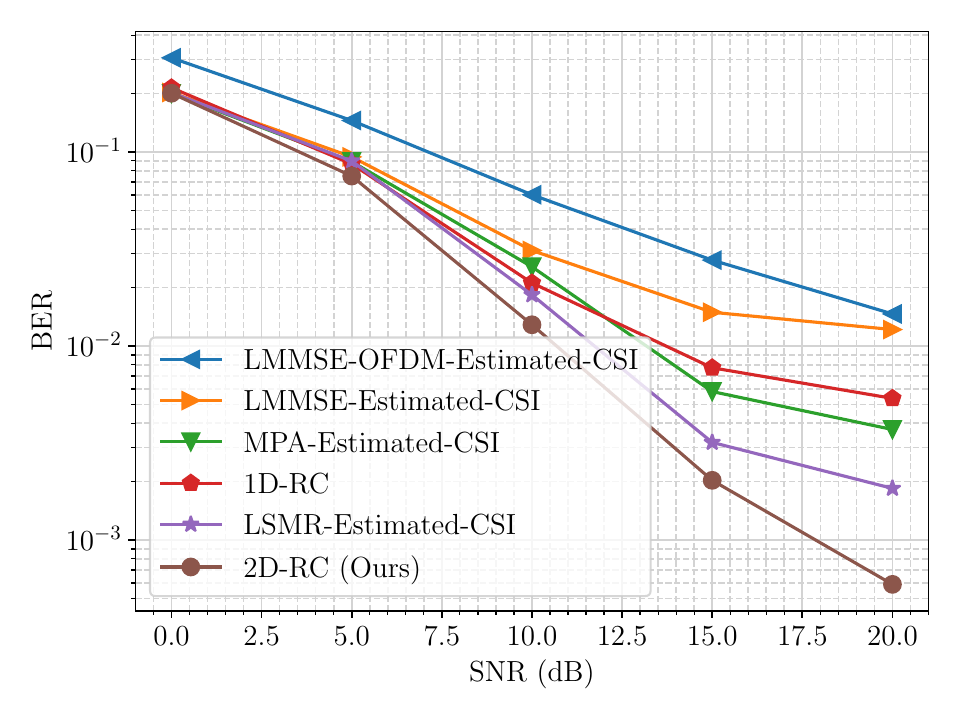}
\caption{BER comparison in the RCP-OTFS system under QPSK.}
\label{figs:ber_rcp_rec_pulse_qpsk_10ns}
\end{figure}
\begin{figure}
\centering
\includegraphics[width=0.8\linewidth]{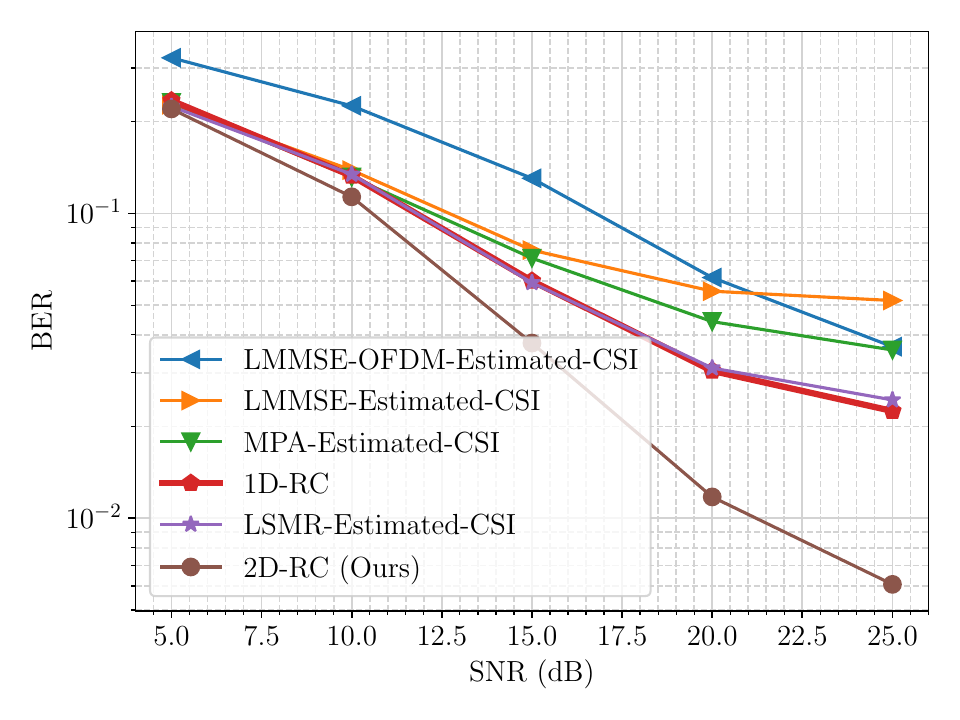}
\caption{BER comparison in the RCP-OTFS system under 16 QAM.}
\label{figs:ber_rcp_rec_pulse_16qam_10ns}
\end{figure}

The following schemes are compared in this paper.
\begin{itemize}
    \item \emph{1D-RC}: The time-domain 1D-RC approach introduced in~\cite{zhou2022learningotfs}, where multiple RCs are required to track the channel changes. 
    The parameters of the 1D-RC approach are set as $N_n=12$, $N_w=10$, and $V=7$, where $V$ denotes the number of 1D-RCs.
    The forget length is searched in the range from $0$ to $22$ with a step size of $2$.
    \item \emph{MPA-Estimated-CSI}: The MPA introduced in~\cite{raviteja2018interference}. 
    The number of iterations is $30$ and the damping factor is set as $0.6$.
    The estimated CSI is obtained by the channel estimation approach in~\cite{raviteja2019embedded}.
    \item \emph{LSMR-Estimated-CSI}: The iterative LSMR-based method~\cite{qu2021TCOMLowcomplexSD} using estimated CSI in~\cite{raviteja2019embedded}. 
    The number of iterations for interference cancellation is set as $5$ and $10$ for QPSK and $16$ QAM, respectively. 
    The number of iterations for LSMR is $15$ for both modulation schemes.
    \item \emph{LMMSE-Estimated-CSI}: The LMMSE detector in the OTFS system, which is implemented in the time domain with the block-wise channel inverse to reduce the computational complexity~\cite{hong2022delay}.
    The CSI is estimated in the DD domain with the approach in~\cite{raviteja2019embedded}.
    \item \emph{LMMSE-OFDM-Estimated-CSI}: The LMMSE equalization in the OFDM system with the LMMSE channel estimation in the TF domain~\cite{hoeher1997two}.
    More details about the adopted pilot pattern in the OFDM system are provided in Appendix~\ref{appendix:pilot_pattern_ofdm}.
\end{itemize}

For the blockwise pilot pattern, pilot symbols are randomly sampled from the modulation alphabet set.
For the spike pilot pattern with guard symbols, the power of the spike pilot is set to ensure that the OTFS subframe with the spike pilot pattern has approximately the same peak-to-average power ratio (PAPR) as utilizing the blockwise pilot pattern.
The reason is that a high PAPR may compel the power amplifier (PA) to operate in the non-linear region, resulting in signal distortion and spectral spreading, as discussed in~\cite{rahmatallah2013peak}.
Therefore, we set the power of the spike pilot by constraining the PAPR.
This setting is equivalent to transmitting the spike pilot with a pilot power of around $20$ dBW for QPSK and $22$ dBW for $16$ QAM.
Depending on the tested signal-to-noise ratio (SNR) and modulation order, the received pilot SNR ranges from around $20$ dB to $47$ dB, which covers the commonly considered pilot SNRs in existing works, e.g.,~\cite{zhou2022learningotfs, liu2021message, thomas2022convolutional}.

\begin{figure}
\centering
\includegraphics[width=0.8\linewidth]{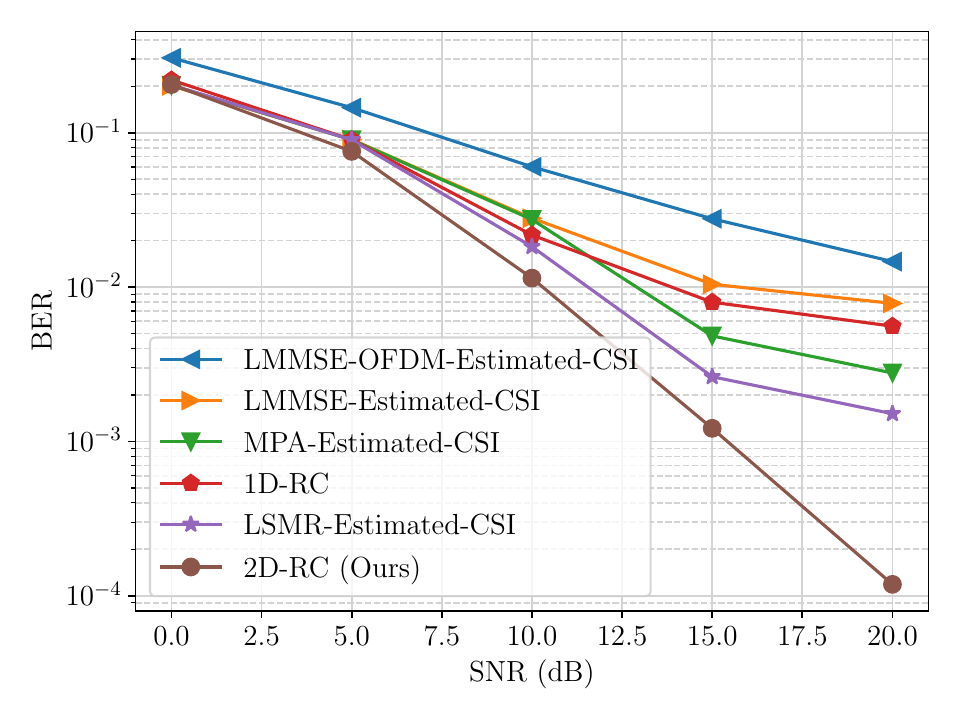}
\caption{BER comparison in the CP-OTFS system under QPSK.}
\label{figs:ber_cp_rec_pulse_qpsk_10ns}
\end{figure}
\begin{figure}
\centering
\includegraphics[width=0.8\linewidth]{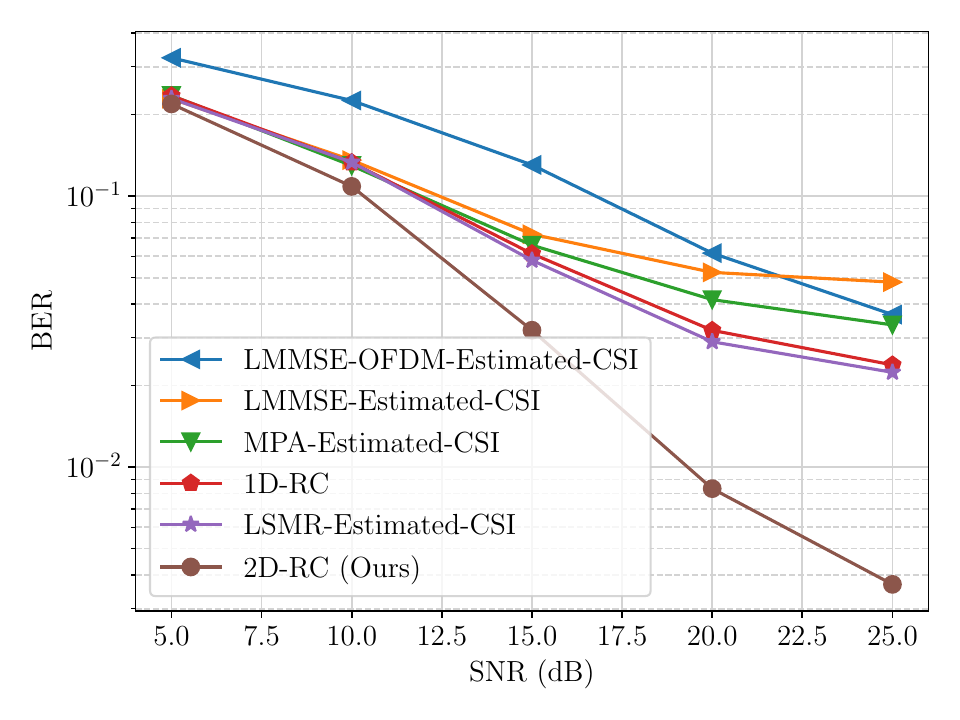}
\caption{BER comparison in the CP-OTFS system under 16 QAM.}
\label{figs:ber_cp_rec_pulse_16qam_10ns}
\end{figure}

In Fig.~\ref{figs:ber_rcp_rec_pulse_qpsk_10ns} and Fig.~\ref{figs:ber_rcp_rec_pulse_16qam_10ns}, we show the bit error rate (BER) performance of different approaches in the RCP-OTFS system under the QPSK and 16 QAM modulations, respectively.
Compared with the existing learning-based 1D-RC method, 2D-RC is demonstrated to have better performance under both the QPSK and 16 QAM modulations, especially in the high SNR regime.
Note that $7$ RCs are utilized in the 1D-RC approach, while only a single NN is exploited for 2D-RC.
The reason is that the 1D-RC method directly adopts the existing RC architecture in the time domain and does not leverage domain knowledge of the OTFS system for its design.
When operating in the time domain, multiple RCs are required to track the changes in the time-varying channel.
Instead, 2D-RC incorporates the 2D circular structure in the DD-domain input-output relationship into its design.
By incorporating structural knowledge, even with a single NN, 2D-RC is more effective than the 1D-RC method that adopts multiple RCs.
The 2D-RC also outperforms compared model-based approaches, i.e., LMMSE, MPA, and the LSMR-based approach, when employing the estimated channel.
Different from the model-based approaches that rely on the knowledge of CSI, the introduced learning-based 2D-RC approach does not require channel knowledge.
Therefore, the performance of 2D-RC is not affected by the accuracy of channel estimates and can be more easily adopted in practical scenarios when it is hard to obtain an accurate CSI.
Furthermore, while a reduced CP overhead is adopted in the RCP-OTFS system, all the considered OTFS-based detectors in the RCP-OTFS system are still shown to perform better than the LMMSE approach in the OFDM system in mid to low SNR regimes.
We further evaluate the performance of compared approaches under both the QPSK and $16$ QAM modulation in the CP-OTFS system.
As illustrated in Fig.~\ref{figs:ber_cp_rec_pulse_qpsk_10ns} and Fig.~\ref{figs:ber_cp_rec_pulse_16qam_10ns}, 2D-RC continues to show an outstanding performance gain over the 1D-RC method and model-based approaches with estimated CSI, which demonstrates the generalization ability of 2D-RC in various scenarios.

\begin{figure}
\centering
\includegraphics[width=0.8\linewidth]{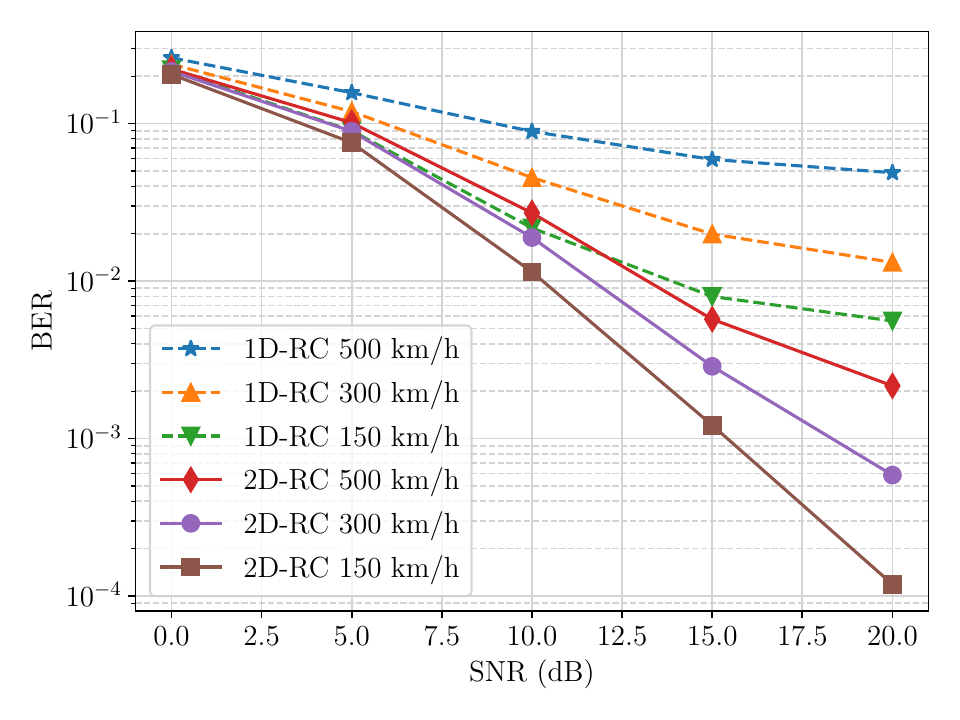}
\caption{BER comparison with different velocities.}
\label{figs:ber_cp_qpsk_10ns_diff_velocities}
\end{figure}

In Fig.~\ref{figs:ber_cp_qpsk_10ns_diff_velocities}, we provide the performance comparison of the 2D-RC and 1D-RC approaches under different user mobility in the CP-OTFS system with QPSK modulation.
As shown in the figure, the 2D-RC approach consistently exhibits a significant performance gain over the 1D-RC method across various velocities, especially in the high SNR regime.
The reason is that the 1D-RC scheme operates in the time domain, where the channel undergoes more substantial changes with the increase of user mobility.
Consequently, as the velocity increases, the disparity between the channel in the pilot region and the channel in the data region becomes more significant.
The mismatch between the training and testing leads to inferior performance of the 1D-RC in higher mobility scenarios.
On the other hand, the 2D-RC incorporates the structural knowledge of the OTFS system into its design and conducts detection in the DD domain.
The increase in mobility causes more severe inter-Doppler interference in the DD domain due to the fractional Doppler effect, resulting in performance degradation of the 2D-RC.
However, the pilot symbols still experience similar channel impairments as the data symbols when the user speed changes.
Due to the reduced training and testing discrepancies, the 2D-RC method demonstrates larger performance gains over the 1D-RC approach in higher mobility cases.

\begin{figure}
\centering
\includegraphics[width=0.8\linewidth]{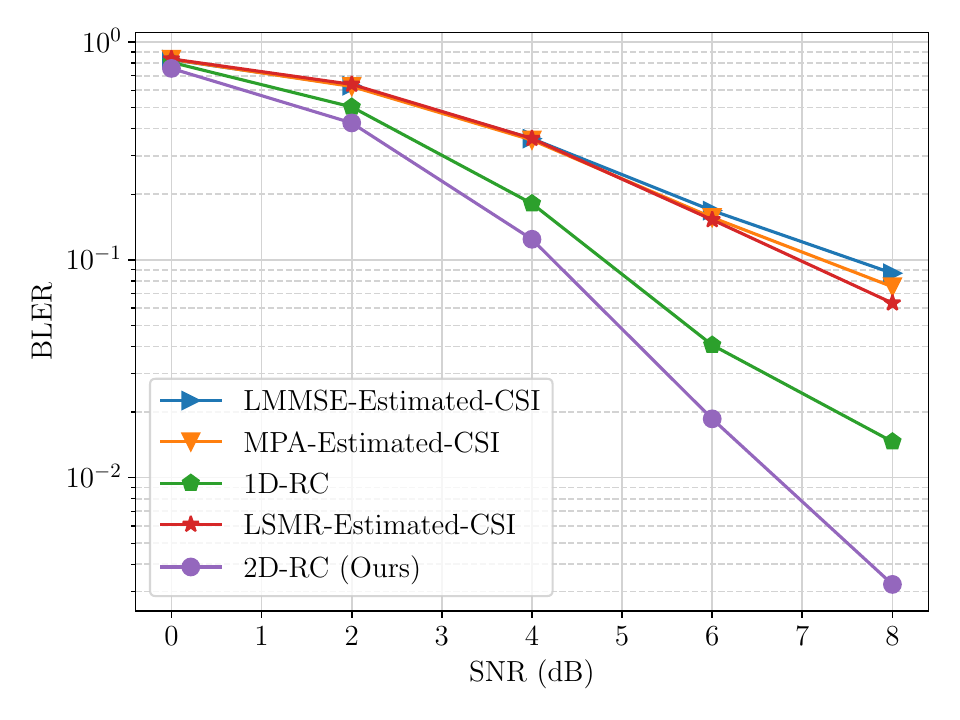}
\caption{BLER comparison with LDPC coding.}
\label{figs:ber_cp_bler_qpsk}
\end{figure}

We also perform the simulation when the low-density parity-check (LDPC) coding is adopted. 
In 3GPP 5G NR~\cite{std3gpp38214}, the code rate can range from 0.0762 to 0.9258. 
In the simulation, the code rate is set as 0.3125.
Fig.~\ref{figs:ber_cp_bler_qpsk} presents the block error rate (BLER) of different approaches in the coded CP-OTFS system with QPSK modulation.
As indicated in the figure, when LDPC coding is exploited, our 2D-RC approach continues to outperform the compared detectors.
Particularly, to achieve a target BLER of $10\%$ specified by the 3GPP 5G NR~\cite{std3gpp38214}, 2D-RC can achieve around 2 dB to 3 dB gain over conventional model-based approaches using estimated CSI. 
The results further demonstrate the effectiveness of the 2D-RC approach when channel coding is utilized.

\begin{figure}
\centering
\includegraphics[width=0.8\linewidth]{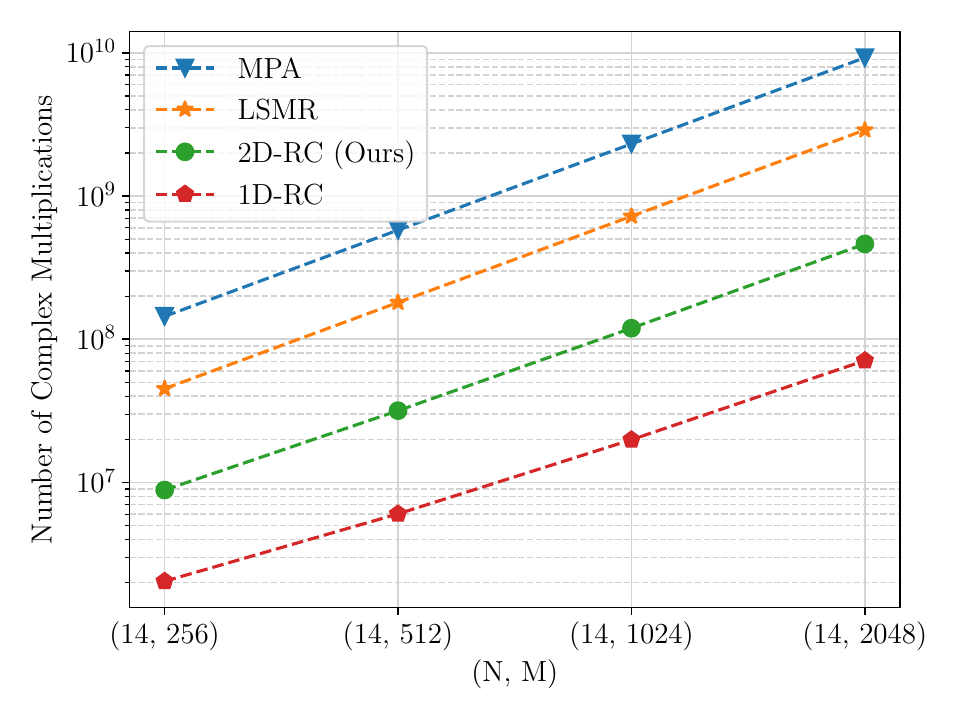}
\caption{Comparison of computational complexity.}
\label{figs:num_complex_mul_16qam}
\end{figure}
Fig.~\ref{figs:num_complex_mul_16qam} shows the total number of complex multiplications for one OTFS subframe as a function of different subframe sizes under 16 QAM modulation.
It is noteworthy that the number of complex multiplications for one OTFS subframe includes both the training\textbackslash{}channel estimation and the testing\textbackslash{}detection for one subframe. 
For model-based approaches such as MPA and LSMR, the computational complexity depends on the number of estimated paths $\tilde{P}$, which includes the number of virtual taps due to the fractional delay and fractional Doppler.
In the evaluation, $\tilde{P}$ is set as the maximum possible number of estimated paths.
Specifically, the maximum possible number of estimated paths is calculated as $\eta MN/2$ when utilizing the channel estimation approach in~\cite{raviteja2019embedded}, where $\eta$ is the pilot overhead.
It can be observed from the figure that the 2D-RC algorithm has lower computational complexity than the model-based MPA and LSMR methods, even when the number of complex multiplications for training is considered in this comparison. 
The low computational complexity includes both training and testing differs 2D-RC from other offline learning methods that rely on a long training time.
While 2D-RC is shown to have higher computational complexity than the 1D-RC method, it can offer a much better performance than the 1D-RC scheme.

\section{Conclusion}
\label{sec:conclusion}

In this paper, we introduce a learning-based 2D-RC approach for the symbol detection task in the OTFS system.
The introduced 2D-RC approach enjoys the same advantage as the previous RC-based approach, which can conduct online subframe-based symbol detection with a limited amount of training data.
The difference is that, unlike the previous RC-based approach that adopts the existing RC structure in the time domain, the introduced 2D-RC scheme is designed to embed the 2D circular channel interaction in the DD domain into its architecture.
By incorporating the domain knowledge of the OTFS system, the 2D-RC approach with a single NN is shown to have significant performance gains over the previous work with multiple RCs in various scenarios.
Furthermore, compared with the model-based approaches, the 2D-RC does not require any channel knowledge and has lower computational complexity.
The results also demonstrate that the 2D-RC outperforms the LMMSE, the MPA, and the LSMR-based method with the estimated CSI across different OTFS system variants and different modulation orders.

\appendices
\section{Input-output relationship with fractional delay and fractional Doppler}
\label{appendix:io_relation}
The vectorized form of the received signal in the DD domain with rectangular pulse shaping can be represented by~\cite{raviteja2018practical}
\begin{align}
    \label{eq:io_relationship_vec_form}
    \mathbf{y} =  (\mathbf{F}_N\otimes\mathbf{I}_M)\mathbf{H}(\mathbf{F}_N^H\otimes\mathbf{I}_M)\mathbf{x},
\end{align}
where $\mathbf{y} = \mathrm{vec}(\mathbf{Y})$ and $\mathbf{x} = \mathrm{vec}(\mathbf{X})$ are the vectorized received and transmitted signal in the DD domain,  
and  $\mathbf{H}\in \mathbb{C}^{MN\times MN}$ is the time-domain channel matrix.

For the RCP-OTFS system, the time-domain channel matrix can be expressed by $\mathbf{H}=\sum_{i=0}^{P-1}h_i \mathbf{\Pi}_{\mathcal{l}_i}\mathbf{\Delta}_{\kappa_i}$,
with $\mathbf{\Pi}_{\mathcal{l}_i} \ \in \mathbb{C}^{MN\times MN}$ models the delay effect of the $i$-th path, and $\mathbf{\Delta}_{\kappa_i} \in \mathbb{C}^{MN\times MN}$ models the Doppler shift effect of the $i$-th path.
Matrices $\mathbf{\Pi}_{\mathcal{l}_i}$ and $\mathbf{\Delta}_{\kappa_i}$ are defined as $\mathbf{\Pi}_{\mathcal{l}_i}\triangleq \mathbf{F}_{MN}\mathbf{D}_{MN}(\mathcal{l}_i)\mathbf{F}_{MN}^H$ and $\mathbf{\Delta}_{\kappa_i}\triangleq \mathbf{D}_{MN}(-\kappa_i)$,
where the  $\mathbf{D}_{MN}(x) \in \mathbb{C}^{MN\times MN}$ is a diagonal matrix with the $(r, c)$-th element $\lbrace\mathbf{D}_{MN}(x)\rbrace_{r,c}=z^{-xr}\delta_{r, c}$; 
the $\delta_{r, c}$ is the Dirac delta function with $\delta_{r, c}=1 ~\text{for} ~ r=c$ and $\delta_{r, c}=0 \text{ otherwise}$.
Define the OTFS modulation matrix as $\mathbf{O} \triangleq \mathbf{F}_N\otimes\mathbf{I}_M \in \mathbb{C}^{MN\times MN}$.
Then \eqref{eq:io_relationship_vec_form} can be written as
\begin{equation}\label{OTFS_eq_ch}
\begin{split}
\mathbf{y}  &= \mathbf{O}\mathbf{H}\mathbf{O}^H\mathbf{x} 
= \sum_{i=0}^{P-1}h_i \underbrace{\mathbf{O}\mathbf{\Pi}_{\mathcal{l}_i}\mathbf{O}^H}_{\triangleq\mathbf{H}_{\mathcal{l}_i}}\underbrace{\mathbf{O}\mathbf{\Delta}_{\kappa_i}\mathbf{O}^H}_{\triangleq\mathbf{H}_{\kappa_i}}\mathbf{x} ,
\end{split}
\end{equation}
where the delay matrix factor for a single path can be further written as $\mathbf{H}_{\mathcal{l}_i} = \mathbf{O} \mathbf{F}_{MN}^H\mathbf{D}_{MN}(\mathcal{l}_i)\mathbf{F}_{MN}\mathbf{O}^H$.
For ease of discussion, we denote $\mathbf{v}_i \triangleq \mathbf{H}_{\kappa_i}\mathbf{x}$ and $\mathbf{y}_i \triangleq \mathbf{H}_{\mathcal{l}_i}\mathbf{v}_i$.
We start by finding an analytical expression for the $(r, c)$-th element of the delay matrix factor $\mathbf{H}_{\mathcal{l}_i}$:
\begin{equation}\label{H_tau1}
\begin{split}
&\lbrace \mathbf{H}_{\mathcal{l}_i}\rbrace_{r,c}=\sum_{t=0}^{MN-1}\lbrace \mathbf{O}\mathbf{F}_{MN}^H  \rbrace_{r,t}\lbrace\mathbf{D}_{MN}(\mathcal{l}_i)\rbrace_{t,t}\lbrace \mathbf{F}_{MN} \mathbf{O}^H \rbrace_{t,c}\\
&=\sum_{t=0}^{MN-1}\lbrace \mathbf{F}_{MN} \mathbf{O}^H \rbrace^*_{t,r}\lbrace\mathbf{D}_{MN}(\mathcal{l}_i)\rbrace_{t,t}\lbrace \mathbf{F}_{MN} \mathbf{O}^H \rbrace_{t,c}\\
&=\frac{1}{M}\sum_{t=0}^{MN-1} z^{t\langle r\rangle_M-t\mathcal{l}_i-t\langle c\rangle_M}\delta_{\langle t\rangle_N, \lfloor \frac{c}{M}\rfloor} \delta_{\langle t\rangle_N, \lfloor \frac{r}{M}\rfloor}.\\
\end{split}
\end{equation}
Let $t=mN+n$ where $m=\lfloor\frac{t}{N}\rfloor$ and $n=\langle t\rangle_N$, then
\begin{equation}\label{H_tau-last}
\begin{split}
&\lbrace \mathbf{H}_{\mathcal{l}_i}\rbrace_{r,c}
=\frac{1}{M}\sum_{m=0}^{M-1}\sum_{n=0}^{N-1} z^{(mN+n)\langle r\rangle_M-(mN+n)\mathcal{l}_i}\\
&\times z^{-(mN+n)\langle c\rangle_M}\delta_{n, \lfloor \frac{c}{M}\rfloor}\delta_{ n, \lfloor \frac{r}{M}\rfloor}\\
&=\frac{1}{M}\sum_{m=0}^{M-1} z^{(mN+\lfloor \frac{c}{M}\rfloor)\langle r\rangle_M-(mN+\lfloor \frac{c}{M}\rfloor)\mathcal{l}_i-(mN+\lfloor \frac{c}{M}\rfloor)\langle c\rangle_M}\\
&\times \delta_{ \lfloor \frac{c}{M}\rfloor, \lfloor \frac{r}{M}\rfloor}\\
&=z^{\lfloor \frac{c}{M}\rfloor (\langle r\rangle_M\! -\! \langle c\rangle_M\! -\! \mathcal{l}_i)}\! \cdot\! \frac{1}{M} \sum_{m=0}^{M-1} z^{m(\langle r\rangle_M\! -\! \langle c\rangle_M\! -\! \mathcal{l}_i)N}\delta_{ \lfloor \frac{c}{M}\rfloor, \lfloor \frac{r}{M}\rfloor}\\\
&=z^{\lfloor \frac{c}{M}\rfloor (\langle r\rangle_M-\langle c\rangle_M-\mathcal{l}_i)} S_M(\langle r\rangle_M-\langle c\rangle_M-\mathcal{l}_i)\delta_{ \lfloor \frac{c}{M}\rfloor , \lfloor \frac{r}{M}\rfloor},
\end{split}
\end{equation}
where $S_M(x) \triangleq \frac{1}{M}e^{j\pi\frac{M-1}{M}x}\frac{\sin \pi  x }{\sin \pi  x/M }$.
Substituting $r=kM+l$ and $c=k'M+l'$ in \eqref{H_tau-last}, we have
\begin{equation}\label{H_tau_2D}
\begin{split}
&\lbrace \mathbf{H}_{\mathcal{l}_i}\rbrace_{kM+l,k'M+l'}
=z^{k (l-l'-\mathcal{l}_i)} S_M(l-l'-\mathcal{l}_i)\delta_{k,k'} \\
&=z^{k (l-l'-\mathcal{l}_i)} \sum_{d=0}^{M-1}\delta_{\langle l-l'\rangle_M, d} S_M(d-\mathcal{l}_i) \delta_{k,k'} \\
&= \sum_{d=0}^{M-1}\alpha_d[l,k]z^{k (\langle l-l'\rangle_M-\mathcal{l}_i)}\delta_{\langle l-l'\rangle_M, d} S_M(d-\mathcal{l}_i) \delta_{k,k'} \\
&= \sum_{d=0}^{M-1}\alpha_d[l,k]z^{k (d-\mathcal{l}_i)}S_M(d-\mathcal{l}_i)\delta_{\langle l-l'\rangle_M, d}  \delta_{k,k'}.
\end{split}
\end{equation}

Denote $V_i[l,k]$ as the $(l, k)$-th element in $\mathbf{V}_i = \text{vec}^{-1}(\mathbf{v}_i)$ and $Y_i[l,k]$ as the $(l, k)$-th element in $\mathbf{Y}_i = \text{vec}^{-1}(\mathbf{y}_i)$.
The $\mathbf{y}_i =\mathbf{H}_{\mathcal{l}_i}\mathbf{v}_i$ is equivalent to
\begin{equation}\label{IO_2D_delay}
\begin{split}
Y_i[l,k]\!&=\!\!\!\sum_{l'=0}^{M-1}\!\!\sum_{d=0}^{M-1}\!\alpha_d[l, k] z^{k (d-\mathcal{l}_i)}S_M(d-\mathcal{l}_i)\delta_{\langle l-l'\rangle_M, d}V_i[l',k']\\
&=\!\!\!\sum_{d=0}^{M-1}\!\alpha_d[l, k] z^{k (d-\mathcal{l}_i)}S_M(d-\mathcal{l}_i)V_i[\langle l-d\rangle_M,k'].
\end{split}
\end{equation}

Similarly, we find an analytical expression for the $(r, c)$-th element of the Doppler matrix factor $\mathbf{H}_{\kappa_i}$:
\begin{equation}
\begin{split}
&\lbrace\mathbf{H}_{\kappa_i}\rbrace_{r,c} \\
&= \frac{1}{N} \sum_{t=0}^{MN-1} z^{r\kappa_i-\lfloor \frac{r}{M}\rfloor \lfloor \frac{t}{M}\rfloor M+\lfloor \frac{t}{M}\rfloor \lfloor \frac{c}{M}\rfloor M} \delta_{\langle r \rangle_M , \langle t\rangle_M } \delta_{\langle t \rangle_M , \langle c\rangle_M }.
\end{split}
\end{equation}		
Let $t=nM+m$ where $m=\langle t\rangle_M$ and $n=\lfloor\frac{t}{M}\rfloor$, then
\begin{equation}\label{H_kappa_i}
\begin{split}
&\lbrace\mathbf{H}_{\kappa_i}\rbrace_{r,c} \\
&=\frac{1}{N} \sum_{n=0}^{N-1}\sum_{m=0}^{M-1} z^{(nM+m)\kappa_i -n \lfloor \frac{r}{M}\rfloor M + n \lfloor \frac{c}{M}\rfloor M} \delta_{\langle r\rangle_M , m} \delta_{m , \langle c\rangle_M }\\
&= z^{\langle c\rangle_M\kappa_i} S_N  (\lfloor \frac{c}{M}\rfloor-\lfloor \frac{r}{M}\rfloor+\kappa_i) \delta_{\langle r\rangle_M , \langle c\rangle_M }.
\end{split}
\end{equation}

Denote $X[l,k]$ as the $(l, k)$-th element in $\mathbf{X} = \text{vec}^{-1}(\mathbf{x})$.
Then $\mathbf{v}_i =\mathbf{H}_{\kappa_i}\mathbf{x}$ is equivalent to
\begin{equation}\label{IO_2D_Doppler}
\begin{split}
V_i[l,k]=\sum_{k'=0}^{N-1}z^{l\kappa_i} S_N(\kappa_i-k')X[l,\langle k-k'\rangle_N].
\end{split}
\end{equation}

Substituting \eqref{IO_2D_delay} and \eqref{IO_2D_Doppler} into \eqref{OTFS_eq_ch} and replacing the variable $d$ with $l'$, we get the final input-output relationship
\begin{equation}
\label{eq:rcp_fractional}
\begin{split}
Y[l,k]&=\sum_{l'=0}^{M-1}\sum_{k'=0}^{N-1}\sum_{i=0}^{P-1}h_i\alpha_{l'}[l, k] z^{k (l'-\mathcal{l}_i)+\kappa_i( \langle l-l'\rangle_M)}
 \\&\times S_M(l'-\mathcal{l}_i)S_N(\kappa_i-k')X[\langle l-l'\rangle_M,\langle k-k'\rangle_N]\\
 &=\sum_{l'=0}^{M-1}\sum_{k'=0}^{N-1}H_{l, k}[l',k']X[\langle l-l'\rangle_M,\langle k-k'\rangle_N],
\end{split}
\end{equation}
where 
\begin{equation}
\label{eq:channel_rcp}
\begin{split}
   &H_{l, k}[l',k']=\sum_{i=0}^{P-1}h_i\alpha_{l'}[l, k] z^{k (l'-\mathcal{l}_i)+\kappa_i( \langle l-l'\rangle_M)}
 \\&\times S_M(l'-\mathcal{l}_i)S_N(\kappa_i-k').
\end{split}
\end{equation}
When $\mathcal{l}_i$ and $\kappa_i$ are integers, the \eqref{eq:rcp_fractional} simplifies to \eqref{eq:rcp_otfs_relationship}.


For the CP-OTFS system, based on the derivation in~\cite{das2020time}, the input-output relationship with fractional delay and fractional Doppler can be written as
\begin{equation}
\label{eq:cp_otfs_fractional}
\begin{split}
Y[l,k]
&=\sum_{l'=0}^{M-1}\sum_{k'=0}^{N-1}\sum_{i=0}^{P-1}h_i\tilde{z}^{\kappa_i(N_{cp}+l-\mathcal{l}_i)}S_M(l-l'-\mathcal{l}_i)\\
&\times S_N(k'-k+\kappa_i)X[l', k'], \\
&=\sum_{l'=0}^{M-1}\sum_{k'=0}^{N-1}\sum_{i=0}^{P-1}h_i\tilde{z}^{\kappa_i(N_{cp}+l-\mathcal{l}_i)}S_M(l' - \mathcal{l}_i)\\
&\times S_N(\kappa_i-k')X[\langle l-l'\rangle_M, \langle k-k'\rangle_N]\\
&=\sum_{l'=0}^{M-1}\sum_{k'=0}^{N-1}H_{l}[l',k']X[\langle l-l'\rangle_M, \langle k-k'\rangle_N].
\end{split}
\end{equation}
where 
\begin{equation}
\label{eq:channel_cp}
\begin{split}
H_{l}[l',k']=\sum_{i=0}^{P-1}h_i\tilde{z}^{\kappa_i(N_{cp}+l-\mathcal{l}_i)}S_M(l' - \mathcal{l}_i)S_N(\kappa_i-k').
\end{split}
\end{equation}
When $\mathcal{l}_i$ and $\kappa_i$ are integers, the \eqref{eq:cp_otfs_fractional} can be written as \eqref{eq:cp_otfs_relationship}.
\section{Pilot pattern in the OFDM system}
\label{appendix:pilot_pattern_ofdm}

\begin{figure}
\centering
\includegraphics[width=0.5\linewidth]{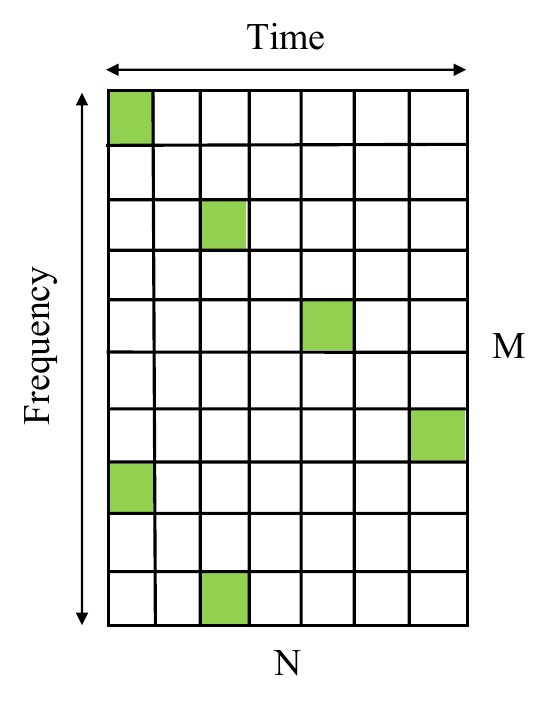}
\caption{Pilot pattern in the OFDM system.
The green grids are filled with known pilot symbols. 
The blank region represents the data symbol position.
}
\label{figs:pilot_pattern_ofdm}
\end{figure}

In the OFDM system, the scattered stairwise pilot pattern in the TF domain is adopted for the LMMSE channel estimation, which is shown in Fig.~\ref{figs:pilot_pattern_ofdm}. 
Specifically, pilots are placed in a scattered way with a spacing of $2$ along both the time and frequency axis to ensure a more accurate channel estimation.
The channel is first estimated at pilot locations and then interpolated over data symbol locations with the channel estimation method in~\cite{hoeher1997two}.
Note that the pilot overhead of this pilot pattern is set to be the same as the pilot patterns utilized in the OTFS system.


\ifCLASSOPTIONcaptionsoff
  \newpage
\fi



\bibliographystyle{IEEEtran}

\bibliography{IEEEabrv,ref.bib}
\end{document}